\documentclass[a4paper,useAMS,usenatbib]{mn2e}
\usepackage[totalwidth=480pt, totalheight=680pt]{geometry}
\usepackage{graphicx}
\usepackage{lastpage}
\usepackage{color}
\def\proptoapprox{\mathrel{\hbox{\rlap{\hbox{\lower4pt\hbox{$\sim$}}}\hbox{$\propto$}}}}
\newdimen\digitwidth
\setbox0=\hbox{2}
\digitwidth=\wd0
\catcode `#=\active
\def#{\kern\digitwidth}

\title[Radio faint AGN: a tale of two populations]
{Radio faint AGN: a tale of two populations}

\author[P. Padovani et al.]{P. Padovani$^{1,2}$\thanks{E-mail:
ppadovan@eso.org}, M. Bonzini$^{1}$, K. I. Kellermann$^{3}$, N. Miller$^{4}$, 
V. Mainieri$^{1}$, P. Tozzi$^{5}$\\
$^{1}$European Southern Observatory, Karl-Schwarzschild-Str. 2,
D-85748 Garching bei M\"unchen, Germany\\
$^{2}$Associated to INAF - Osservatorio Astronomico di Roma, via Frascati 33, I-00040
Monteporzio Catone, Italy\\
$^{3}$National Radio Astronomy Observatory, 520 Edgemont Road, 
 Charlottesville, VA 22903-2475, USA\\
$^{4}$Department of Mathematics and Physical Sciences, Stevenson University, 
1525 Greenspring Valley Road, Stevenson, MD  21153-0641, USA\\
$^{5}$INAF - Osservatorio Astrofisico di Arcetri, Largo E. Fermi, I-50125, Firenze, Italy\\
}

\begin{document}

\date{Accepted ... Received ...; in original form ...}

\pagerange{\pageref{firstpage}--\pageref{lastpage}} \pubyear{2015}

\maketitle

\label{firstpage}

\begin{abstract}
  We study the Extended {\it Chandra} Deep Field South (E-CDFS) Very Large
  Array sample, which reaches a flux density limit at 1.4 GHz of $32.5~\mu$Jy
  at the field centre and redshift $\sim 4$, and covers $\sim 0.3$
  deg$^2$. Number counts are presented for the whole sample while the
  evolutionary properties and luminosity functions are derived for active
  galactic nuclei (AGN). The faint radio sky contains two totally distinct
  AGN populations, characterised by very different evolutions, luminosity
  functions, and Eddington ratios: radio-quiet (RQ)/radiative-mode, and
  radio-loud/jet-mode AGN. The radio power of RQ AGN evolves
  $\proptoapprox (1+z)^{2.5}$, similarly to star-forming galaxies, while the
  number density of radio-loud ones has a peak at $z \sim 0.5$ and then
  declines at higher redshifts. The number density of radio-selected RQ AGN
  is consistent with that of X-ray selected AGN, which shows that we are
  sampling the same population. The unbiased fraction of radiative-mode RL
  AGN, derived from our own and previously published data, is a strong function 
  of radio power, decreasing from $\sim 0.5$ at
  $P_{\rm 1.4GHz} \sim 10^{24}$ W Hz$^{-1}$ to $\sim 0.04$ at $P_{\rm 1.4GHz}
  \sim 10^{22}$ W Hz$^{-1}$. Thanks to our enlarged sample, which now
  includes $\sim 700$ radio sources, we also confirm and strengthen our
  previous results on the source population of the faint radio sky:
  star-forming galaxies start to dominate the radio sky only below $\sim 0.1$
  mJy, which is also where radio-quiet AGN overtake radio-loud ones.
\end{abstract}


%

\section{Introduction}\label{intro}

Soon after the discovery of quasars \citep{sch63} it was realized that the
majority of them were not as strong radio sources as the first quasars, as
they were undetected by the radio telescopes of the time: they were
``radio-quiet'' \citep{san65}. It was later understood that these sources
were actually only ``radio-faint'', as for the same optical power their radio
powers were $\approx 3 - 4$ orders of magnitude smaller than their radio-loud
(RL) counterparts, but the name stuck. Radio-quiet (RQ) active galactic
nuclei (AGN) were until recently normally found in optically
selected samples and are characterized by relatively low radio-to-optical
flux density ratios ($R \la 10$) and radio powers ($P_{\rm 1.4GHz} \la
10^{24}$ W Hz$^{-1}$ locally).

Innumerable studies have compared the properties of the two AGN classes in
various bands to try and shed light on their inherent differences. As a
result, the distinction between the two types of AGN has turned out to be not
simply a matter of semantics: the two classes represent intrinsically
different objects, with RL AGN emitting a large fraction of their energy
 non-thermally and in association with powerful
relativistic jets, while the multi-wavelength emission of RQ AGN is dominated
by thermal emission, directly or indirectly related to the accretion
disk. The host galaxies are also different, with those of RL AGN being
elliptical while those of RQ ones, excluding the most powerful ones, being
spiral \citep[e.g.][]{dun03}.

However, fifty years after the discovery of quasars the question ``Why do only a
minority of galaxies that contain an AGN have jets?'' is still unanswered. One
problem is that, while it would be important to compare the properties of the
two AGN classes in the band where they differ most (i.e. the radio band) this
requires the identification of AGN in deep ($\la 1$ mJy) radio fields, as RQ AGN
are, by definition, radio faint. This has been possible only recently
\citep[e.g.][and references therein]{pad09,pad11b,Sim12,white15}.

Our group started addressing this topic, together with others including the
broader issue of the source population in very deep radio fields, initially
in the {\it Chandra} Deep Field South (CDFS). This was done by defining a
complete sample of 198 radio sources sources reaching $\sim 43~\mu$Jy over
0.2 deg$^2$ at 1.4 GHz using the National Radio Astronomy Observatory (NRAO)
Very Large Array (VLA) through a multi-wavelength approach
\citep{kel08,mai08,toz09,pad09,pad11b}.

\cite{mil08,mil13} expanded on this by observing the so-called Extended CDFS
(E-CDFS), again using the VLA, down to $\sim 30~\mu$Jy at $5\sigma$, in a
2.8" x 1.6" beam over $\sim 0.3$ deg$^2$. This resulted in a sample of almost
900 sources. We have started exploiting these new radio data with the aim of
addressing the issues of the faint radio source population and RQ AGN in more
detail, given the larger and slightly deeper E-CDFS sample, as compared to
the CDFS one. In particular, \cite{bon12} have identified the optical and
infrared (IR) counterparts of the E-CDFS sources, finding reliable matches
and redshifts for $\sim 95\%$ and $\sim 81\%$ of them respectively, while
\cite{vat12} have identified the X-ray counterparts and studied the radio --
X-ray correlation for star forming galaxies (SFG). Finally, \cite{bon13}
have provided reliable source classification and studied the host galaxy 
properties of RQ AGN, RL AGN, and SFG.

The main aims of this paper are:

\begin{enumerate}

\item present the most accurate number counts for the various classes of
  sub-mJy sources down to $\sim 30~\mu$Jy;

\item study the evolution and luminosity functions (LFs) of sub-mJy AGN. In
  particular, our CDFS paper on this topic \citep{pad11b} was affected by
  small number statistics, particularly for RQ AGN, and by the large fraction
  ($\sim 50\%$) of upper limits on the far-IR-to-radio flux density ratio
  ($q$) used for classifying sources. As discussed at length in
  \cite{pad11b}, this topic has important implications for constraining the
  mechanism behind radio emission in RQ AGN and allow a proper comparison
  with RL AGN;

\item investigate more deeply the density evolution of RL AGN found in \cite{pad11b}
($\propto (1+z)^{-1.8}$);

\item provide a new approach to the issue of the fraction of RL sources
  within the AGN population.

\end{enumerate}

Items (i) and (ii) are very relevant, together with the study of the
evolution of SFG, also to the issue of the link between the black holes at
the centre of AGN and their host galaxies. Moreover, they are also critical
to predict the source population at radio flux densities $<1~\mu$Jy, which
are important, for example, for the Square Kilometre Array (SKA). All
existing estimates, in fact, had to rely, for obvious reasons, on
extrapolations and are based on high flux density samples. This affects
particularly the highest redshifts, which can better be probed at fainter
flux densities.

Our physical definition of the various classes of sub-mJy sources follows
that of \cite{pad11b}, to which we refer the reader. Following \cite{hec14}
we will also use the terms ``radiative-mode'' and ``jet-mode''
AGN\footnote{See Table 4 of \cite{hec14}.}. In short, radiative-mode RL AGN
include radio quasars and high-excitation radio galaxies, while jet-mode RL
AGN refer to low-excitation radio galaxies. There is considerable overlap
between jet-mode AGN and the radio sources morphologically classified as
Fanaroff-Riley type I (FR I) \citep{fan74} and also between RL radiative-mode
AGN and FR IIs, although there is a sizeable population of jet-mode FR IIs
and a smaller one of radiative-mode FR I \citep[e.g.][and references
therein]{gen13}. As for RQ AGN, radiative-mode sources are the ``classical''
broad- and narrow-lined AGN (type I and II), while the jet-mode ones are the
so-called liners. Note that the two classes have also widely different
Eddington ratios, with radiative-mode and jet-mode sources respectively above
and below $L/L_{\rm Edd} \approx 0.01$.

In this paper we present the general properties (number counts, redshift
distribution, etc.) of the whole sample and then concentrate on the evolution
and LFs of AGN. We will discuss in detail SFG in a future publication
(Padovani et al., in preparation).

Section~2 describes the VLA-E-CDFS sample, while Section~3 presents the so-far
deepest determination of sub-mJy number counts by class. Section~4 studies
the sample evolution while Section~5 derives the AGN LFs and their evolution. 
Finally, Section~6 discusses our results, while section~7
summarizes our conclusions. Throughout this paper spectral indices are
defined by $S_{\nu} \propto \nu^{-\alpha}$ and the values $H_0 = 70$ km
s$^{-1}$ Mpc$^{-1}$, $\Omega_{\rm M} = 0.3$, and $\Omega_{\rm \Lambda} = 0.7$
have been used.

\section{The sample}\label{sample}

\subsection{Redshifts}\label{Sect_redshift}

Our original sample is described in \cite{bon12} and \cite{mil13} and includes
883 radio sources detected in 0.324 deg$^2$ at 1.4 GHz in a deep VLA survey of
the E-CDFS. The average $5\sigma$ flux density limit is $\sim 37~\mu$Jy,
reaching $\sim 30~ \mu$Jy in the field centre. The fraction of sources with
redshift information is $\sim 81\%$, which is not ideal for determining the
evolutionary properties and LFs of our sample. We have then excluded the
outermost part of the field, where we do not have enough ancillary
multi-wavelength data to provide reasonable photometric redshifts. The resulting
sub-sample covers an area of 0.285 deg$^2$ \citep[see][for details]{bon13} and
includes 765 radio sources with radio flux density $\ge 32.5~\mu$Jy, 87\% of
which have redshifts (of these, $\sim 40\%$ are spectroscopic and $\sim 60\%$
are photometric). This is the sample we use to build the number counts in
Sect. \ref{sec_number_counts}.
To further increase the fraction of sources with redshift, we have made
another cut and considered only sources with a $3.6\mu$m flux density
$f_{3.6\mu m} >$ 1 microJy, which is approximately the completeness limit of
the IRAC observations at $3.6\mu$m (only four sources have $f_{3.6\mu m} <$ 1
microJy). Coupled with a radio flux density limit of $32.5~\mu$Jy, this
provides us with clean selection criteria and with a sample of 680 radio
sources, 92\% of which have redshift \citep[the same fraction as
  in][]{pad11b}.
We estimate the redshift for the $\sim 8\%$ of the objects in the sample
without observed redshift from their $f_{3.6\mu m}$, as detailed in Appendix
\ref{sec:red_estim}. To further minimise the effects of these redshift
estimates we also define two subsamples: AGN (and RQ AGN) with
$z\le3.66$ and SFG with $z\le3.25$.  Above these two values, in fact, 80\%
and 95\% of the redshifts, respectively, are estimated. By imposing such
cuts, we are then only assuming that the excluded sources are at high
redshifts, which makes sense based on their $f_{3.6\mu m}$ values (see
Appendix \ref{sec:red_estim}), without actually using their values.  
This is the sample we use mostly to study the evolution and LF of AGN in 
Sect. \ref{evolution} and \ref{sect:LF}. 

Note that the effect of these redshift estimates on our results is minimal,
for two reasons: 1. our final redshift incompleteness is very small ($\sim
3\%$, $\sim 5\%$, and $\sim 1.5\%$ for all AGN, RQ, and RL AGN respectively:
see Sect. \ref{VeVa_evolution}); 2. redshift affects $V_{\rm e}/V_{\rm a}$
values (Sect. \ref{VeVa_evolution}) much less than flux density.
Nevertheless, error bars on the LFs are evaluated using the number of sources
per bin with redshift determination only and when binning in redshift the
percentages of estimated redshifts are also given for each bin.

\subsection{Classification}\label{class}

Source classification has been done by \cite{bon13}, expanding upon the
scheme of \citet{pad11b}. In short, by using $q_{24obs}$, that is the
logarithm of the ratio between the observed 24$~\mu$m and 1.4 GHz flux
densities, we could define an ``SFGs locus'' based on the radio -- FIR
correlation for SFG. Sources below this locus display a radio excess that is
the signature of an AGN contribution to the radio luminosity and were
therefore classified as RL AGN. Within and above this locus, a source was
classified as a RQ AGN if there was evidence of AGN activity in the other
bands considered. Namely, if it had a hard band ($2-10$ keV) X-ray luminosity
greater than $10^{42}$ erg s$^{-1}$ or it lay within the ``AGN wedge'' of the
IRAC colour-colour diagram, as defined by \cite{donley12}. Otherwise, the
object was classified as an SFG. We point out that our AGN classification
includes all sources independently of optical appearance (i.e., quasar-like
and galaxy-like). We refer the reader to \cite{bon13} for
further details. 

The issue of the classification of our sources into radiative- or jet-mode
sources has been briefly discussed by \cite{bon13}. The optimal approach to
be able to make such a distinction requires rest-frame optical/UV emission
line ratios but, given the relatively high redshift of our sample and the
incomplete spectroscopic coverage, this is only possible for a handful of
objects. \cite{bon13}, using a criterion based on the 22 $\mu$m
power proposed by \cite{gur14}, estimated that the large majority of our RL
sources are of the jet-mode type; this is confirmed by their LF
(Sect. \ref{rlagn:LF_local}). As regards our RQ AGN, given our selection
criteria, which include also a cut in X-ray power, we expect them to be for
the large part of the radiative-mode type. This issue is addressed in 
Sect. \ref{astro}.

We stress that ours is the deepest sample for which both the evolution and LFs 
of radio sources has been studied. 

\section{The sub-mJy number counts}\label{sec_number_counts}

\begin{figure*}
\includegraphics[width=0.49\textwidth]{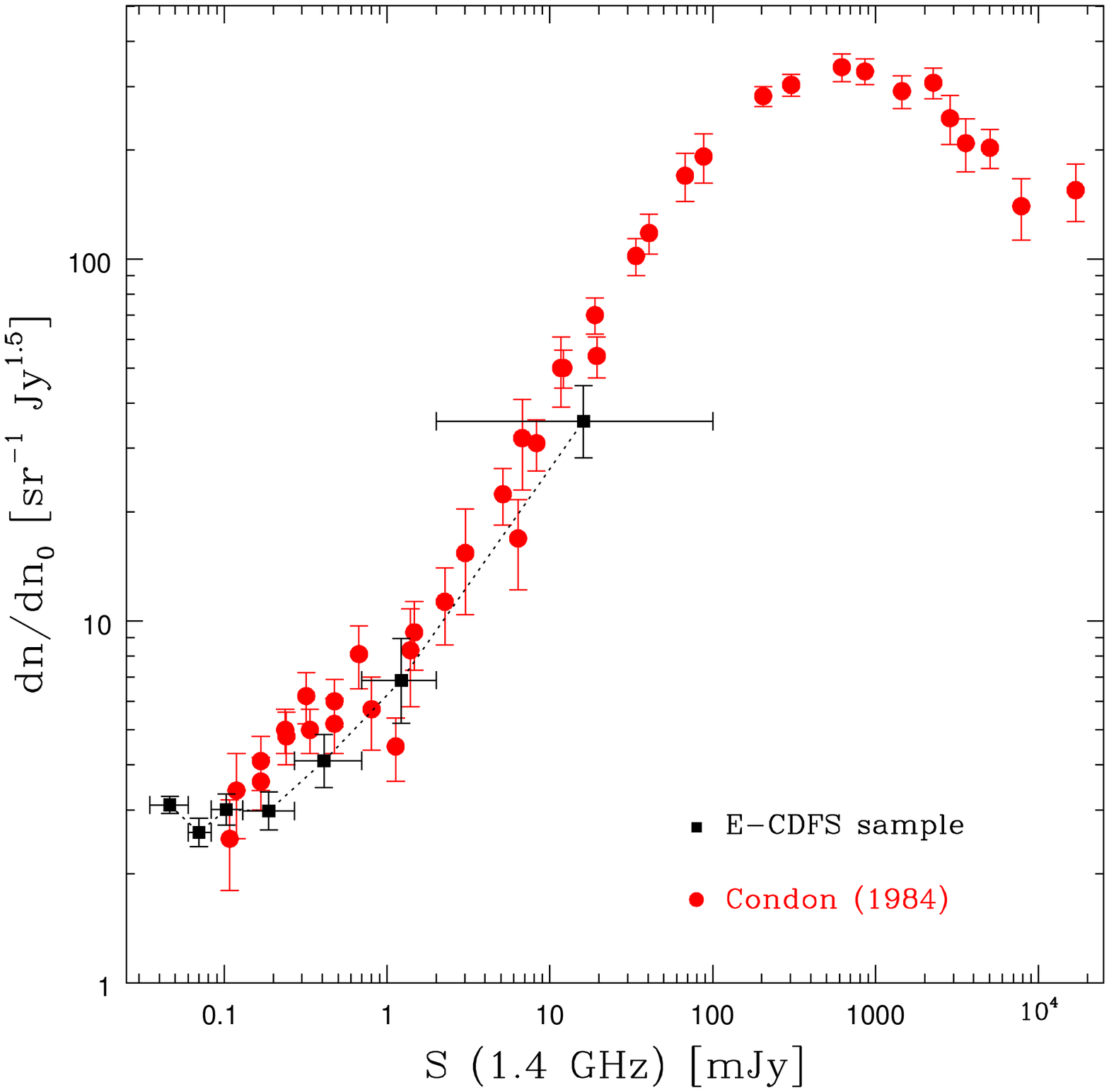}
\includegraphics[width=0.49\textwidth]{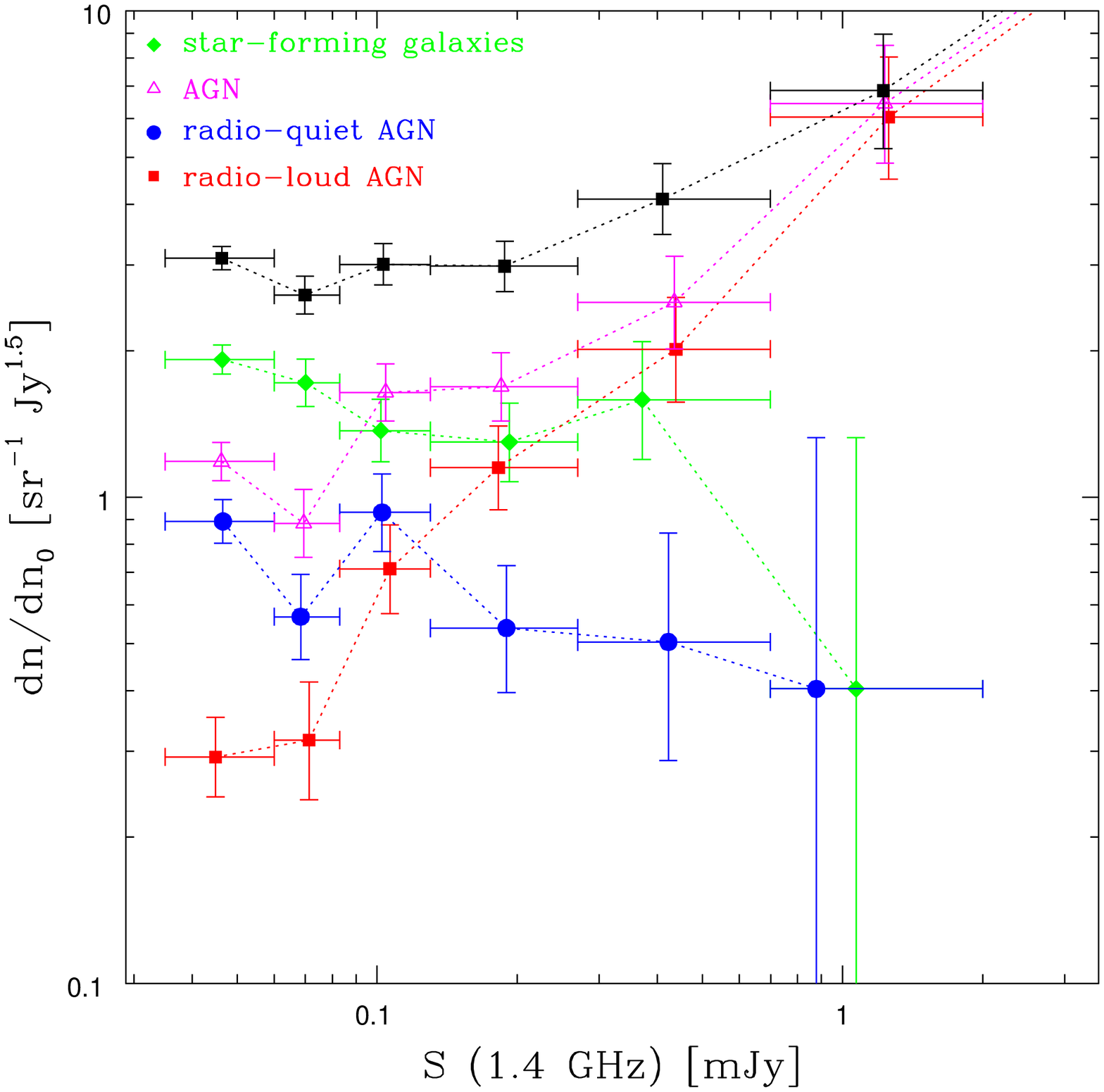}
\caption{a) (left) the Euclidean normalised 1.4 GHz ECDFS source counts (black
  filled squares) compared to the source counts at 1.4 GHz from the compilation
  of \protect\cite{con84} (red filled circles); b) (right) the Euclidean
  normalised 1.4 GHz ECDFS source counts for the whole sample (black filled
  squares) and the various classes of radio sources: SFGs (green diamonds), all
  AGN (magenta triangles), RQ AGN (blue circles), and RL AGN (red
  squares). Error bars correspond to $1\sigma$ Poisson errors \citep{geh86}. See
  text for details.}
\label{counts_classes}
\end{figure*}

\begin{table*}
\caption{Euclidean normalised 1.4 GHz counts. \label{tab_counts}}
\begin{tabular}{lcccccccc}
$f_{\rm r}$ range & mean $f_{\rm r}$ &   & & & Counts &  & \\
                 & & ##Total & SFG & Fraction & AGN & Fraction & RL AGN & RQ AGN \\
###($\mu$Jy) & ($\mu$Jy) & (sr$^{-1}$ Jy$^{1.5}$) & (sr$^{-1}$ Jy$^{1.5}$) & (\%) &
  (sr$^{-1}$ Jy$^{1.5}$) & (\%) &
  (sr$^{-1}$ Jy$^{1.5}$) &
  (sr$^{-1}$ Jy$^{1.5}$)\\
\hline
$#32.5 - 60$ & #46 &$3.10^{+0.17}_{-0.16}$ & $1.92^{+0.14}_{-0.13}$ & $62^{+6}_{-5}$#
& $1.18^{+0.11}_{-0.10}$ & $38^{+4}_{-4}$# & $0.29^{+0.06}_{-0.05}$ & $0.89^{+0.10}_{-0.09}$\\
$##60 - 83$ & #70 &$2.60^{+0.25}_{-0.23}$ & $1.72^{+0.20}_{-0.18}$ & $66^{+10}_{-9}$ 
& $0.88^{+0.15}_{-0.13}$ & $34^{+6}_{-6}$# & $0.32^{+0.10}_{-0.08}$ & $0.57^{+0.13}_{-0.10}$\\
$##83 - 130$ & 103 &$3.01^{+0.31}_{-0.28}$ & $1.37^{+0.22}_{-0.19}$ & $46^{+8}_{-8}$# 
& $1.64^{+0.24}_{-0.21}$ & $54^{+9}_{-9}$# & $0.71^{+0.17}_{-0.14}$  & $0.93^{+0.19}_{-0.16}$\\
$#130 - 270$ & 188 & $2.99^{+0.38}_{-0.34}$ & $1.30^{+0.26}_{-0.22}$ & $43^{+10}_{-9}$ 
& $1.69^{+0.29}_{-0.25}$ & $57^{+12}_{-11}$ & $ 1.15^{+0.25}_{-0.21}$  & $0.54^{+0.19}_{-0.14}$\\
$#270 - 700$ & 410 & $4.11^{+0.75}_{-0.64}$ & $1.59^{+0.50}_{-0.39}$ & $39^{+14}_{-12}$ 
& $2.52^{+0.61}_{-0.50}$ & $61^{+18}_{-17}$ & $ 2.01^{+0.56}_{-0.45}$  & $0.50^{+0.34}_{-0.22}$\\
$#700 - 2000$ & 1223 & $6.86^{+2.10}_{-1.65}$ & $0.40^{+0.92}_{-0.33}$ & $6^{+13}_{-5}$ 
& $6.45^{+2.05}_{-1.59}$ & $94^{+37}_{-37}$ & $6.05^{+2.00}_{-1.54}$  & $0.40^{+0.92}_{-0.33}$\\
$2000 - 100000$ & 16092 &$ 35.6^{+9.1}_{-7.4}$ & ... & ... ## & $35.6^{+9.1}_{-7.4}$ & $100^{+33}_{-33}$ 
& $35.6^{+9.1}_{-7.4}$  & ... \\
\hline
\end{tabular}
\end{table*}
   
More than half a century ago, the earliest radio source counts gave the first
indications of cosmic evolution \citep{Ryle55} but the results were not without
controversy. The central problem in understanding source counts is the conflict
between source confusion at lower resolution and missing lower surface
brightness sources or underestimating their flux density at high
resolution. These issues, while better understood, still exist today.  There is
no ``correct'' resolution to use that avoids the effects of confusion and
resolution, but there is an optimum resolution depending on the areal density of
sources. Modern deep radio surveys reach flux density levels of only a few
$\mu$Jy where the corresponding source density is $\sim 5$ sources arcmin$^{-2}$
at $S= 10~\mu$Jy \citep{vernstrom_2015}. So a resolution of at least a few
arc seconds is needed to unambiguously separate sources. But many $\mu$Jy
sources are typically an arcsecond in size, so are partially resolved with a one
or two arcsecond beam \citep[e.g.][]{mux05}.

Published deep radio source counts are widely discrepant, even when made with
the same instrument, and even with different investigators using the same
instrument in the same field \citep{con12}. Since source detection depends on
the peak flux density, the effects of resolution cause two problems. First, if a
source is resolved the measured peak flux density needs to be multiplied by the
ratio of integrated to peak flux density. For strong sources, where both the
measured and peak flux densities can be accurately determined this is
straightforward. But for sources with lower signal-to-noise ratio (SNR) Gaussian
fitting routines give systematically spurious large sizes and a corresponding
flux density bias \citep[e.g.][]{con97}. Even more complicated are sources where
the peak flux density is below the detection limit, but with integrated flux
densities which are above the detection limit. To account for these
non-detections one normally may use a source size distribution equivalent to
that of stronger survey sources, but this does not necessarily give the correct
answer. As described by \cite{mil13} we have used peak flux densities, except
where we measure a significantly higher integrated value, a procedure which
depends on the local SNR. Because of the then limitations of the VLA at the time
of the \cite{mil13} survey, the 25 MHz IF bands could be divided into only seven
sub bands each of 3.125 MHz which resulted in significant bandwidth smoothing
away from the field centre. Because each source may appear in more than one
subfield field, the corrections for bandwidth smoothing are not straightforward
\citep{mil13}.

An additional problem arises when one physical source is split into two or
more distinct components.  Unless there is a connection between the radio
features, there is no way to a-priori distinguish between a relatively strong
multi-component source or multiple weaker sources. But the derived source
count will depend dramatically on which interpretation is used.  We have
addressed this issue by reference to optical and/or X-ray counterparts using
a maximum likelihood criteria to decide which optical counterparts are
associated with which radio features \citep{bon12}.

Finally, we note the problem caused by the so-called Eddington bias. Because
the source counts are steep, that is there are more weak sources than strong
sources, random noise fluctuations cause more weak sources to appear above
any flux density value, than strong sources which fall below this value. As
described by \cite{wall12} the determination of the Eddington bias to the
observed source count is not straightforward. 
Following the procedure described by 
\cite{wall12}, Wall (private communication) has calculated the Eddington bias corrections 
to the source count assuming a Euclidean slope. We then
corrected the number of sources by a factor, which depended on SNR and
ranged from 1/1.194 ($5 \le SNR < 6$) to 1/1.052 ($9 \le SNR < 10$). These same
factors were also applied when deriving the LFs.

To investigate the relative contribution of the different source types to the
radio population as a function of flux density, we consider here the
sub-sample within an area of 0.285 deg$^2$, which provides us with more
information for this purpose (i.e., we do not make any $3.6\mu$m flux cut). 
The sensitivity of our survey is a function of
the position in the field of view, although a much less strong one than for
the single pointing CDFS sample. In more detail, the area covered is equal
to 0.285 deg$^2$ and independent of flux density above $\sim 60~\mu$Jy,
decreases quite slowly to 0.279 deg$^2$ and 0.226 deg$^2$ at $\sim 50~\mu$Jy
and $\sim 40~\mu$Jy respectively, and reaches 0.05 deg$^2$ at the flux
density limit \citep[see][]{mil13}. We included sources down to a root mean square 
(rms) noise of 6.5 $\mu$Jy, that is a flux density limit of 32.5 $\mu$Jy. 

Fig. \ref{counts_classes}a compares the E-CDFS number count with previous
determinations at relatively high flux densities compiled by \cite{con84}. The
strong source count above $\sim 1$ Jy shows the familiar steep slope resulting
from the well know dramatic evolution of radio loud quasars and radio galaxies.
Below this value, all of the strongest sources in Universe have already been
included, and the count begins to converge. Below 1 mJy, the count steepens
(flattens in the normalised plot)
again due to the emergence of the low luminosity population of RQ AGN and SFG,
which are discussed further in this paper and in the next one of this series.

Table \ref{tab_counts} and Fig. \ref{counts_classes}b\footnote{A
  preliminary version of this figure has been shown by \cite{pad14}.} present
the Euclidean normalised number counts for the various classes. The
normalised number counts agree very well in the region of overlap (flux
densities $\ge 43$ $\mu$Jy) with those of the CDFS \citep{kel08}\footnote{Due
  to a small numerical error, the number counts in \cite{kel08} and,
  consequently, also those in \cite{pad09} and \cite{pad11b}, were too high
  by $\sim 5\%$.} and appear to be slightly higher at fainter flux densities. The
comparison with the number counts of \cite{pad11b} is not shown for
clarity but the two determinations agree very well, typically within $\sim 1
\sigma$. The main results of \cite{pad11b} are also confirmed: AGN dominate
at large flux densities ($\ga 1$ mJy) but SFGs become the dominant population
below $\approx 0.1$ mJy. Similarly, RL AGN are the predominant type above 0.1
mJy but their counts drop fast at lower values.
 
In more detail, AGN make up $43\pm3\%$ \citep[where the errors are based on
binomial statistics:][]{geh86} of sub-mJy sources and their counts are seen
to drop at lower flux densities, going from 100\% of the total at $\sim 10$
mJy down to $38\%$ at the survey limit. SFG, on the other hand, which
represent $57\pm4\%$ of the sample, are missing at high flux densities but
become the dominant population below $\approx 0.1$ mJy, reaching $62\%$ at
the survey limit. RQ AGN represent $26\pm2\%$ (or $61\%$ of all AGN) of
sub-mJy sources but their fraction appears to increase at lower flux
densities, where they make up $75\%$ of all AGN and $\approx 29\%$ of all
sources at the survey limit, up from $\approx 6\%$ respectively at $\approx
1$ mJy. The relative fractions of the various classes in our sample are shown
in Fig. 6 of \cite{bon13}.

\section{Evolution}\label{evolution}

\subsection{$V_{\rm e}/V_{\rm a}$ analysis}\label{VeVa_evolution}

We first study the evolutionary properties of the VLA-E-CDFS sample through a
variation of the $V/V_{\rm max}$ test \citep{sch68}, the $V_{\rm e}/V_{\rm
  a}$ test \citep{av80, mor91}, that is the ratio between {\it enclosed} and
{\it available} volume, since we do not have a single flux density limit (see
Sect. \ref{sec_number_counts}). More details on this and other statistical
techniques used in this paper can be found in \cite{pad11b}.

We have computed $V_{\rm e}/V_{\rm a}$ values for our sources taking into
account our double flux density limits and the appropriate sky coverage
\citep[see eqs. 42 and 43 of][]{av80}. The IR k-correction was done by using
the measured near-IR spectral indices. In the radio band spectral indices
between 1.4 and 4.86 GHz are available for $\sim 29\%$ of the sources
\citep{kel08,huy12}, while for the rest the mean values for the relevant
classes were assumed. Statistical errors are given by $\sigma =
1/\sqrt{12~N}$ \citep{av80}. We estimate the significance of the deviation
from the non-evolutionary case by deriving the $p$-value that 
the $V_{\rm e}/V_{\rm a}$ distribution is similar to a uniform one according to
a Kolmogorov-Smirnov (KS) test. Similar results are obtained by using the
deviation from 0.5 (the value expected for no evolution) of $\langle V_{\rm
  e}/V_{\rm a} \rangle$. To have an initial simple estimate of the sample
evolution we have also derived the best fit parameter $k_{L}$ assuming a pure
luminosity evolution (PLE) of the type $P(z) = P(0) (1+z)^{k_L}$ or a pure
density evolution (PDE) of the type $\Phi(z) = \Phi(0) (1+z)^{k_D}$, where
$\Phi(z)$ is the luminosity function.

\begin{figure}
\includegraphics[height=8.6cm]{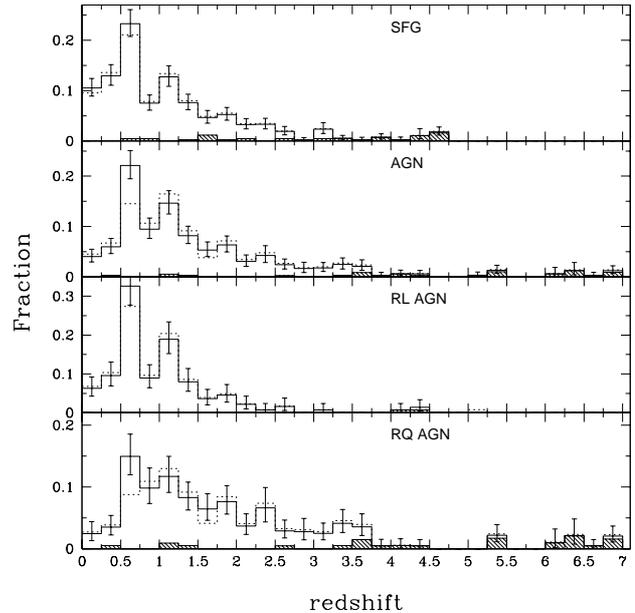}
\vspace{0.8cm}
\caption{Fractional redshift distributions for the different classes of
  sources, deconvolved with the appropriate sky coverage. The dashed areas
  denote redshifts estimated from the $3.6\mu$m flux density. Error bars represent
  the $1\sigma$ range based on Poisson statistics. The dotted lines take into
  account the presence of LSS as discussed in Appendix \ref{sec:LSS}.}
\label{histz}
\end{figure}

\begin{table*}
\caption{Sample Evolutionary Properties: $V_{\rm e}/V_{\rm
a}$ analysis.}
\begin{tabular}{lrcrcrrr}
Sample & N & $\langle z \rangle$ & est. $z$ \% & $\langle V_{\rm e}/V_{\rm
a}\rangle$ & $p$-value#& $k_L$$^{a}$ & $k_D$$^{b}$\\
\hline
All sources & 680 & $1.35\pm 0.05$ & 8.4\%& $0.609\pm0.011$ &  $<0.001$ & 
$1.6\pm0.1$ & ...  ##\\
Star-forming galaxies & 372 & $1.16\pm 0.05$ & 8.9\% &$0.656\pm0.015$ & $<0.001$ &
$2.1\pm0.1$& ...  ##\\
Star-forming galaxies, $z \le 3.25$ & 356 & $1.01\pm 0.04$ & 5.1\% &$0.651\pm0.015$ & $<0.001$ &
$2.4\pm0.2$ & ...  ##\\
All AGN & 308 & $1.54\pm 0.08$ & 7.8\% &$0.550\pm0.016$ & $<0.001$ &
$0.9\pm0.2$& ... ##\\
All AGN, $z \le 3.66$ & 288 & $1.27\pm 0.05$ & 2.8\% & $0.552\pm0.017$ & 0.003 & 
$1.1\pm0.3$& ... ##\\
All AGN, $z \le 3.66$, no LSS & 264 & $1.31\pm 0.05$ & 3.0\% & $0.556\pm0.018$ & 0.003 & 
$1.2\pm0.3$& ... ##\\
RQ AGN & 172 & $1. 93\pm 0.11$& 12.8\%& $0.709\pm0.022$ & $<0.001$ &
#$2.3\pm0.1$ & ...  ##\\
RQ AGN, $z \le 3.66$ & 155 & $1. 54\pm 0.07$& 5.2\%& $0.708\pm0.023$ & $<0.001$ &
#$3.0\pm0.2$ & ...  ##\\
RQ AGN, $z \le 3.66$, no LSS & 141 & $1. 61\pm 0.08$& 5.7\%& $0.711\pm0.024$ & $<0.001$ &
#$3.0\pm0.2$ & ...  ##\\
RL AGN & 136& $1.00\pm 0.06$& 1.5\%& $0.354\pm0.025$ & $<0.001$ &
$-3.9\pm0.9$ & $-2.1\pm0.3$\\
RL AGN, no LSS & 126& $1.03\pm 0.07$& 1.6\%& $0.362\pm0.026$ & $<0.001$ &
$-3.6\pm0.9$ & $-2.0\pm0.3$\\

\hline
\multicolumn{5}{l}{\footnotesize $^{a}$Pure luminosity evolution $P(z) = P(0)(1+z)^{k_L}$}\\
\multicolumn{5}{l}{\footnotesize $^{b}$Pure density evolution $\Phi(z) = \Phi(0) (1+z)^{k_D}$}\\
\end{tabular}
\label{tabveva}
\end{table*}

Our results are shown in Table \ref{tabveva}, which gives the sample in
column (1), the number of sources in column (2), the mean redshift in column
(3), the percentage of sources with redshift estimated from the magnitude in
column (4), $\langle V_{\rm e}/V_{\rm a} \rangle$ in column (5), the $p$-value 
that the $V_{\rm e}/V_{\rm a}$ distribution is similar to a 
uniform one in column (6), and the best fit parameters $k_L$ and
$k_D$ (when applicable) in column (7) and (8) respectively. When the best fit
indicates negative luminosity evolution (i.e., $k_{L} <0$), in fact, we fit a
PDE model as well, which we feel is more physical in this
case. The mean redshift
is calculated taking into account the effect of the sky coverage, that is
each object is weighted by the inverse of the area accessible at the flux
density of the source \citep[e.g.][]{pad07}. 

The fractional redshift distributions for the different classes are shown in
Fig. \ref{histz}. As also shown in Table \ref{tabveva}, RQ AGN have the
highest $\langle z \rangle$ ($z \sim 2$) and reach $z \sim 6 - 7$
(photometric redshifts). RL AGN have also a relatively broad distribution,
extending up to $z \sim 4.5$ but with a lower $\langle z \rangle$ than RQ
ones ($z \sim 1$). SFG are in between but closer to RL AGN. The main results
on the sample evolution are the following:

\begin{enumerate}
\item The whole sample has $\langle V_{\rm e}/V_{\rm a} \rangle > 0.5$ and
  shows significant evolution characterized by $k_L=1.6\pm0.1$;
  
\item SFG evolve at a very high significance level ($p$-value $<0.001$);
  their evolutionary parameter for the case of pure luminosity evolution is
  $k_L=2.1\pm0.1$. This increases to $k_L=2.4\pm0.2$ for a redshift cut $z
  \le 3.25$, which reduces the fraction of estimated redshifts. The
  evolution and LF of SFG will be presented and discussed in Padovani et al., 
  in preparation;
  
\item AGN as a whole evolve relatively weakly, with $k_L=0.9\pm0.2$
  ($k_L=1.1\pm0.3$ for $z \le 3.66$);
     
\item RQ AGN, however, evolve very significantly ($p$-value $<0.001$) with
  $k_L=2.3\pm0.1$, consistent with the value of SFG (a thorough comparison
  between the two classes is deferred to Padovani et al., in preparation). This
  increases to $k_L=3.0\pm0.2$ for a redshift cut $z \le 3.66$, which reduces
  the fraction of estimated redshifts;
   
\item RL AGN also evolve strongly ($p$-value $<0.001$) but in the
  negative sense, with $k_L=-3.9\pm0.9$ and $k_D=-2.1\pm0.3$. This evolution
  is in the opposite sense as that of RQ AGN. As we show below
  (Sect. \ref{banded} and \ref{rlagn:LF_z}), the best-fit evolution of RL AGN
  is more complicated than assumed here.
 \end{enumerate}
 
We note that \cite{pad11b} found a difference in $\langle V_{\rm e}/V_{\rm a}
\rangle$ between RL AGN with powers below and above $P = 10^{24.5}$ W Hz$^{-1}$,
with the former having $\langle V_{\rm e}/V_{\rm a} \rangle \approx 0.5$ and the
latter reaching smaller values. Given the flux density-limited nature of our
sample, and the resulting degeneracy between power and redshift, we believe that
this difference, which is still present in the current sample, is better
characterised as a redshift dependence (see Sect. \ref{rlagn:LF_z}).
  
Given the small area of our survey one could worry that the presence of
large-scale structures (LSS) \citep[e.g.][]{gil03} and related redshift
spikes might influence some of our results. The LSS in the E-CDFS is
discussed in Appendix \ref{sec:LSS}. To assess the impact of the LSS on our
study, most of our results are presented for the ``full'' and the ``no
LSS'' sub-samples, as illustrated, for example, in Fig. \ref{histz} (cf. the
solid with the dotted lines). As shown in Table \ref{tabveva}, the resulting
$\langle V_{\rm e}/V_{\rm a} \rangle$ values and best-fit evolutionary
parameters for the latter are well within $1 \sigma$ from those derived for
the former, which shows that the effect of the LSS on these results is
minimal. These over-densities are obviously more noticeable when one studies
the evolution of the LF with redshift (see, e.g. the top-middle and
bottom-left panels of Fig. \ref{rqagn_z_bins}) but even then the revised LFs
are typically within $\la 1\sigma$ from the ``full'' ones.

\subsection{The banded $\langle V_{\rm e}/V_{\rm a} \rangle$}\label{banded}

\begin{figure}
\includegraphics[height=8.6cm]{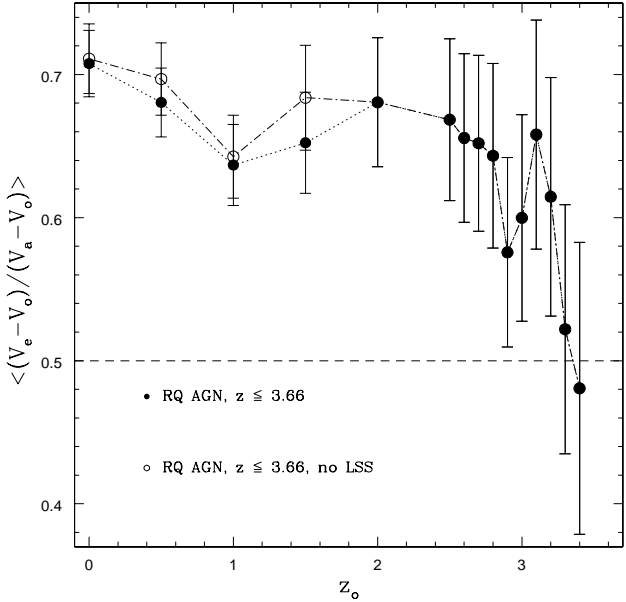}
\vspace{0.8cm}
\caption{The banded statistic, $\langle (V_{\rm e} - V_o)/(V_{\rm a} - V_o)
  \rangle$ versus $z_o$, for our RQ AGN with $z \le 3.66$. The horizontal
  dashed line indicates the value of 0.5, expected under the null hypothesis
  of no evolution. Open symbols take into account the presence of LSS as
  discussed in Appendix \ref{sec:LSS}.}
\label{fig.vovm_z}
\end{figure}

Given the available statistics, we can easily study possible changes of the
evolution with redshift by using the so-called banded $\langle V/V_{\rm max}
\rangle$ statistic, i.e. the quantity $\langle (V - V_{\rm o})/(V_{\rm max} -
V_{\rm o}) \rangle$, or $\langle (V_{\rm e} - V_{\rm o})/(V_{\rm a} - V_{\rm
  o}) \rangle$ in our case, where $V_{\rm o}$ is the cosmological volume
enclosed by a given redshift $z_o$ \citep{Dunlop90}. This procedure allows us
to detect any high-redshift (possibly negative) evolution by separating it by
the well-known strong (positive) low-redshift evolution.

\begin{figure}
\includegraphics[height=8.6cm]{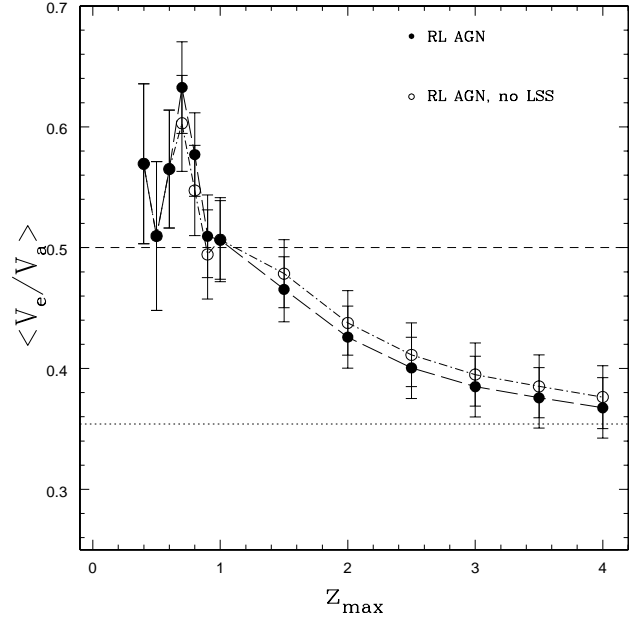}
\vspace{0.8cm}
\caption{$\langle V_{\rm e}/V_{\rm a} \rangle$ as a function of maximum
  redshift for our RL AGN. The horizontal dashed line indicates the value of
  0.5, expected under the null hypothesis of no evolution, while the dotted
  line is the value for the sample. Open symbols take into account the
  presence of LSS as discussed in Appendix \ref{sec:LSS}.}
\label{fig.vovm_zmax}
\end{figure}

Fig. \ref{fig.vovm_z} shows that $\langle (V_{\rm e} - V_{\rm o})/(V_{\rm a}
- V_{\rm o}) \rangle$ for RQ AGN has a very slow decline with redshift, but
still with values significantly $> 0.5$, until it reaches 0.5 at $z \approx
3.3$. This suggests that the evolution might be somewhat weaker at higher
redshifts; we will address this in Sects. \ref{rqagn:LF_z} and
\ref{sect:rqagn_evol}.

As for the RL AGN $\langle V_{\rm e}/V_{\rm a} \rangle$ is $< 0.5$ for the
whole sample so the banded $\langle V/V_{\rm max} \rangle$ statistic is not
useful. However, we can plot $\langle V_{\rm e}/V_{\rm a} \rangle$
vs. the maximum redshift of the sub-sample to see if there is evidence of
positive or at least null evolution at low redshifts. This is done in
Fig. \ref{fig.vovm_zmax}, which shows that, indeed, up to $z \approx 1.5$
$\langle V_{\rm e}/V_{\rm a} \rangle$ is not significantly $<0.5$, being
actually $> 0.5$ at the $\sim 2 - 3 \sigma$ level at $z \sim 0.7 -
0.8$. While part of this is due to the presence of LSS, taking these into
account still results in $\langle V_{\rm e}/V_{\rm a} \rangle$ $\sim 0.5$ and
actually $> 0.5$ at the $2.6\sigma$ level at $z \sim 0.7$. At $z > 1$
$\langle V_{\rm e}/V_{\rm a} \rangle$ declines with redshift, reaching
gradually its sample value. This shows that the RL AGN evolution changes from
being positive/null to being negative around $z \approx 1 - 1.5$ and that
therefore it cannot be characterized by a simple $(1+z)^k$ law. In other
words, $\langle V_{\rm e}/V_{\rm a} \rangle$ evaluated for the whole sample
tells only part of the story.  We will expand on this in
Sect. \ref{rlagn:LF_z}.

\subsection{Maximum Likelihood Analysis}\label{sect:maxl}

A more general approach to estimate the evolution, and at the same time the
LF, is to perform a maximum likelihood fit of an evolving luminosity function
to the observed distribution in luminosity and redshift. This approach makes
maximal use of the data and is free from arbitrary binning; however, unlike
the $V_{\rm e}/V_{\rm a}$ test, it is model dependent. We consider one and
two power-law LFs, that is $\Phi(P) \propto P^{-\gamma_1}$ and $\Phi(P)
\propto 1/[(P/P_*)^{\gamma_1} + (P/P_*)^{\gamma_2}]$ respectively, with a
break at $P_*$. We refer the reader to \cite{pad11b} for more details.

Our results are shown in Table \ref{tabmaxl}, which gives the sample in
column (1), the evolutionary model in column (2), the two LF slopes (if
applicable) in columns (3) and (4), the best-fit evolutionary parameter in
column (5), and the break power (if applicable) in column (6). Errors are
$1\sigma$ for one interesting parameter. The best-fit evolutionary parameters
agree (within $1 - 2\sigma$) with those derived through the $V_{\rm
  e}/V_{\rm a}$ approach.

The maximum likelihood approach (but also the $V_{\rm e}/V_{\rm a}$ one) does
not provide a goodness-of-fit test. Therefore, as it is typically done, we
carry out a KS test on both the cumulative redshift and radio power
distributions and compute the probability of having the KS test statistics as
large or larger than the observed one. This can be used to reject a model
when too low. The $p$-values are given in columns (7) (KS$_{\rm z}$) and (8)
(KS$_{\rm P}$) respectively, where the $> 2\sigma$ and $> 3\sigma$ levels
correspond to values $< 0.046$ and $< 0.003$ respectively. The fact that in
most cases one of the two $p$-values is $\le 0.04$ shows that the models we
adopted are too simplistic, as detailed in the next Section.

\begin{table*}
\caption{Sample Luminosity Functions and Evolution: maximum likelihood analysis
  best fits. We consider a LF of the type $\Phi(P) \propto
    1/[(P/P_*)^{\gamma_1} + (P/P_*)^{\gamma_2}]$ and an evolution $\propto
    (1+z)^{k}$.  KS$_{\rm z}$ and KS$_{\rm P}$ are the KS $p$-values for the
    redshift and radio power distributions respectively.}
\begin{tabular}{llccrcrr}
Sample&Model&$\gamma_1$&$\gamma_2$&$k$&$\log P_*$&KS$_{\rm z}$&KS$_{\rm P}$\\
& & & & &{W~Hz$^{-1}$}& & \\
\hline
AGN, $z \le 3.66$  & PDE & $1.70\pm0.04$ & ... & $0.11\pm0.26$ & ... & $<$0.001& 0.016 \\
AGN, $z \le 3.66$  & PLE & $1.70\pm0.04$  & ... & $0.16^{+0.33}_{-0.41}$ & ... & $<$0.001& 0.016\\
AGN,  $z \le 3.66$  & PDE & $1.2^{+0.1}_{-0.2}$ & $1.82\pm0.06$ &
$0.0\pm0.3$ & $22.6\pm0.6$& 0.006 & 0.306 \\
AGN,  $z \le 3.66$  & PLE & $1.2\pm0.2$ & $1.80\pm0.05$ &
$-0.15\pm0.35$ & $22.6\pm0.6$& 0.005 & 0.340 \\
AGN,  $z \le 3.66$, no LSS & PLE & $1.3\pm0.2$ & $1.80^{+0.10}_{-0.05}$ &
$0.1\pm0.4$ & $22.5^{+1.2}_{-0.9}$& 0.039 & 0.383 \\
RQ AGN,  $z \le 3.66$  & PLE  & $2.3\pm0.1$ & ...  &
$2.8\pm0.2$ & ... & 0.022 & 0.038\ \\
{\bf RQ AGN}, $\mathbf{z \le 3.66}$  & {\bf PLE }& $0.6^{+0.6}_{-0.7}$ & $2.64^{+0.19}_{-0.27}$ &
$2.52^{+0.20}_{-0.23}$ & $22.32^{+0.20}_{-0.16} $& 0.372 & 0.885 \\
{\bf RQ AGN,}  $\mathbf{z \le 3.66}$, {\bf no LSS} & {\bf PLE} & $0.9^{+0.5}_{-0.6}$ & $2.72^{+0.25}_{-0.23}$ &
$2.54\pm0.23$ & $22.50\pm0.25 $& 0.471 & 0.879 \\
RL AGN & PLE & $1.40^{+0.05}_{-0.04}$ & ... & $-6.0\pm1.4$ & ... & 0.003& 0.771 \\
RL AGN & PDE & $1.40\pm0.04$ & ...  & $-2.4\pm0.3$ & ... & 0.003 & 0.772 \\
RL AGN, no LSS & PDE  & $1.41\pm0.04$ & ... & $-2.3\pm0.3$ & ... & 0.018& 0.870\\
\hline
\multicolumn{8}{l}{\footnotesize Models consistent with the data are indicated in boldface.}
\end{tabular}
\label{tabmaxl}
\end{table*}

\section{Luminosity Functions}\label{sect:LF}

The results of the maximum likelihood approach expand upon and complement those
of the $V_{\rm e}/V_{\rm a}$ test. A simple power law fit for the LF of AGN is
excluded with very high significance ($p$-value $<0.001$), both in the case of PLE
and PDE. However, even a double power law fit is inconsistent with the data,
which means that the adopted evolutionary laws are too simple. A simple power
law fit for the LF of RQ AGN is also excluded ($p$-value $\sim 0.02$), while a PLE
evolution plus a two power-law LF is consistent with the data. As for RL AGN,
neither a simple PLE or PDE model can fit the data, while a two power-law LF is
not needed (the best fit slopes turn out to be the same well within the
errors). The two AGN classes, apart from displaying evolutions with different
signs, have also widely different LF shapes, with RL AGN being characterised by
a much flatter LF. Note that the effect of the LSS on these results is truly
minimal ($< 1\sigma$).

We now turn to study the AGN LFs in more detail. We first study all AGN together but 
then, given their widely different LFs and evolutions, we examine the two classes separately.

The effect of cosmic variance was included in the LF uncertainties as follows: 1. we first
estimated $\sigma_{\rm dm}$ for the relevant redshift range following
\cite{moster_2011} (using the IDL code provided in their paper); 2. we then
derived the bias $b$ for the appropriate redshift and number density from Fig. 3
(left) of \cite{somerville_2004}. We could not use in fact 
    \cite{moster_2011} to get $b$ as well given that their numbers apply to the whole galaxy
    population above a given stellar mass, while we here deal with a radio-selected sample; 3. we estimated the relative cosmic variance
as $\sigma_v = b \times \sigma_{\rm dm}$ \citep[see][for details and definitions]
{moster_2011};
4. $\sigma_v$ was then added in quadrature with the Poisson uncertainties
for each sub-population in each redshift bin. We note
that, since the number of objects per power bin in our LFs is never very
large, the former turned out always to be smaller than the latter.

\begin{figure}
\includegraphics[height=8.6cm]{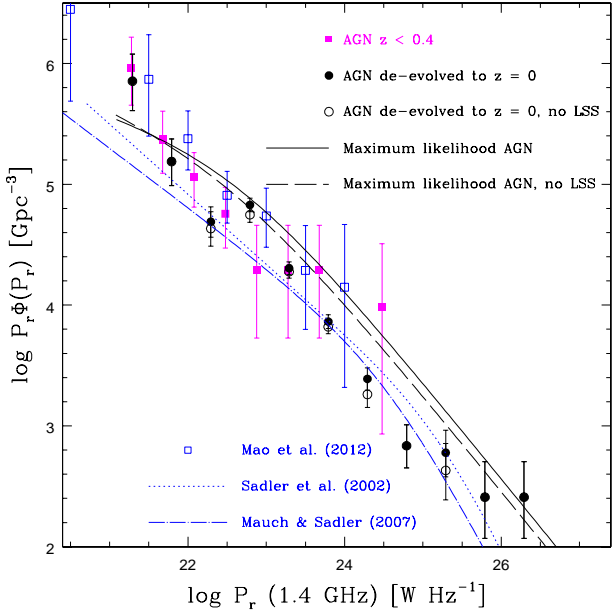}
\caption{The local differential 1.4 GHz LF for AGN in a $P \times \phi(P)$
  form. Filled (magenta) squares indicate the E-CDFS LF at $z \le 0.4$, while
  filled (black) circles denote the LF de-evolved to $z=0$ using the best fit
  evolutionary parameter from the $\langle V_{\rm e}/V_{\rm a} \rangle$
  analysis, with open circles taking into account the presence of LSS. The
  solid line is the best-fit double power-law LF from the maximum likelihood
  method (PLE), while the dashed line is the same LF but taking into account
  the presence of LSS. Open (blue) squares refer to the $z < 0.2$ ATLAS LF
  \citep{Mao12}. The best fits to the local AGN LF from \protect\cite{sad02}
  (converted to our value of $H_0$) and \protect\cite{mau07} are also shown
  (blue dotted and dash-dotted lines respectively). Error bars for the E-CDFS  
  $z \le 0.4$ sample correspond to
  $1\sigma$ Poisson errors evaluated using the number of sources per bin with
  redshift determination added in quadrature with the cosmic variance
  uncertainties.}
\label{agn_local}
\end{figure}

\subsection{All AGN}\label{sec:All_AGN_LF}

Fig. \ref{agn_local} shows three different estimates of the local LF for our
AGN, done using the $1/V_{\rm max}$, or in our case $1/V_{\rm a}$, method.  The
AGN $z \le 0.4$ sample, which includes 29 objects, is consistent with no
evolution ($\langle V_{\rm e}/V_{\rm a} \rangle = 0.57\pm0.05$) and can
therefore be used as a robust proxy for the local AGN LF (filled squares;
similar results are obtained by using the $z \le 0.3$ sample, which however
contains only 21 objects). To increase our statistics, we have included in the
sample all sources above our radio completeness limit independently of their
$f_{3.6\mu m}$ values (see Sect. \ref{Sect_redshift}), which adds four more
sources\footnote{We do not need in fact to restrict ourselves to sources with
  $f_{3.6\mu m} >$ 1 microJy to increase redshift completeness, since none of
  the AGN with $z < 0.4$ have redshift estimated from $f_{3.6\mu m}$.}. In
Fig. \ref{agn_local} we also plot the best fits to the local ($z \le 0.3$) LFs
from \cite{sad02} and \cite{mau07} (dotted and dash-dotted lines
respectively). Our local LF is consistent within the errors ($\chi^2_{\nu} \la 1.5$) with both but 
nominally a factor $\sim 1.7 - 2.3$ higher than the
\cite{sad02} and \cite{mau07} LF respectively.

Our $z < 0.4$ LF agrees with the $z < 0.2$ AGN LF of \cite{Mao12}, based on
the Australia Telescope Large Area Survey (ATLAS) at 1.4 GHz, which currently
reaches $\sim 150 \mu$Jy per beam. The \cite{smo09} AGN LF in the $0.1 -
0.35$ redshift range (not shown for clarity) is also consistent
with ours.

Fig. \ref{agn_local} shows two further estimates of the local LF for E-CDFS
AGN, which are based on the whole sample but are model dependent. The first one
is the LF de-evolved to $z=0$ (filled circles) using the best fit evolutionary
parameter from the $\langle V_{\rm e}/V_{\rm a} \rangle$ analysis. This LF is
not consistent ($\chi^2_{\nu} \sim 6.1 - 8.2$ respectively, $p$-value $< 0.0001$)
with both previous LFs, which indicates that the redshift evolution is more
complex than assumed. The situation does not change taking into account the LSS
($\chi^2_{\nu} \sim 5.0 - 6.2$, $p$-value $< 0.0001$). The second one is the LF
derived from the maximum likelihood analysis (double power-law fit, PLE),
including also the LSS effect, which is $\sim 2 - 3$ times above previous
determinations.

This situation is somewhat at variance with that in \cite{pad11b}, where the
local AGN LF was different ($p$-value $\sim 0.011$) from both the \cite{sad02} and
\cite{mau07} LF. We attribute this difference to a statistical fluctuation
due to the smaller (15 sources vs. 29) local AGN sample available in that
paper. As regards the fact that our LFs are still somewhat above previous
LFs, we noted in \cite{pad11b} that our AGN LF includes the contribution of
RQ AGN, which were not present in significant numbers in these two LFs as
these were based on the NRAO VLA Sky Survey ($S_{\rm min} \ga 2.8$ mJy),
while RQ AGN make up a non-negligible fraction of radio sources only below
$\approx 1$ mJy (Fig. \ref{counts_classes}). Moreover, both previous LFs
included only non-stellar optical sources (i.e., they excluded quasars),
while we have no such cut. The issue of the RQ AGN contribution can be dealt
with by studying RL AGN only.

\subsection{RL AGN}

\subsubsection{The local LF}\label{rlagn:LF_local}

\begin{figure}
\includegraphics[height=8.6cm]{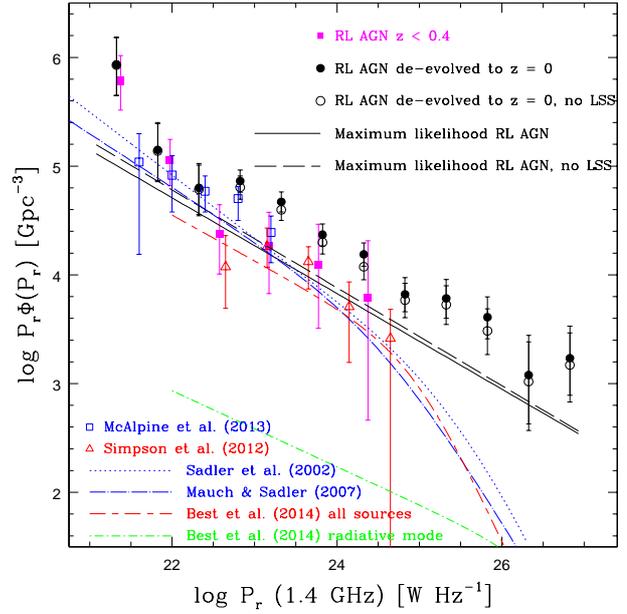}
\caption{The local differential 1.4 GHz LF for RL AGN in a $P \times \phi(P)$
  form. Filled (magenta) squares indicate the E-CDFS LF at $z \le 0.4$, while
  filled (black) circles denote the LF de-evolved to $z=0$ using the best fit
  evolutionary parameter from the $\langle V_{\rm e}/V_{\rm a} \rangle$
  analysis, with open circles taking into account the presence of LSS. The
  solid line is the best-fit from the maximum likelihood method (ZDE),
  while the dashed line is the same LF but taking into account the presence
  of LSS. Open (blue) squares refer to the $0.05 < z < 0.35$ VIDEO-{\it XXM3}
  LF \citep{McAl13}, while open (red) triangles are the $ z \le 0.5$
  Subaru/{\it XMM-Newton Deep Field} LF \citep{Sim12}, corrected to take into
  account the different selection criteria as described in the text. The best
  fits to the local AGN LF from \protect\cite{sad02} (converted to our value
  of $H_0$), \protect\cite{mau07} (blue dotted and dash-dotted lines
  respectively) and \protect\cite{best14} (all sources: red short-long dashed
  line; radiative-mode: green dot - short-dashed line) are also shown. Error
  bars for the E-CDFS  
  $z \le 0.4$ sample correspond to $1\sigma$ Poisson errors evaluated using the number of
  sources per bin with redshift determination added in quadrature with the cosmic variance uncertainties.}
\label{rl_agn_local}
\end{figure}

Fig. \ref{rl_agn_local} shows three different estimates of the local LF for
our RL AGN.  The RL AGN $z \le 0.4$ sample, which includes 23 objects, is
consistent with no evolution ($\langle V_{\rm e}/V_{\rm a} \rangle =
0.57\pm0.06$) and can therefore be used as a robust proxy for the local RL
AGN LF (filled squares). As done for the full AGN sample, this includes all
sources above our radio completeness limit independently of their $f_{3.6\mu
  m}$ values (see Sect. \ref{Sect_redshift}). Our local LF is consistent
within the errors ($\chi^2_{\nu} \la 1.3$) with both the LFs from \cite{sad02}
and \cite{mau07} (dotted and dash-dotted lines respectively), although still
a factor $\sim 1.4 - 1.8$ higher.

We also compare our local LF with the $ z \le 0.5$ Subaru/{\it XMM-Newton
  Deep Field} LF of \cite{Sim12} (red open triangles). That paper adopts a
somewhat different method to classify AGN into RL and RQ sources, based on
the rest-frame ratio of mid-IR to radio flux density $q_{24} = \log(S_{\rm
  24\mu m}/S_{\rm 1.4GHz})$, with RL AGN defined by $q_{24} \le -0.23$.  Our
method relies instead on the observed $q_{24}$ and the dividing line is
redshift dependent. To take this difference into account, we checked how many
RL AGN with $z \le 0.5$ and flux density $\ge 100~\mu$Jy \citep[the limit
of][]{Sim12} were classified as RL AGN by our method and by that of
\cite{Sim12} and corrected their LF by the ratio of the two (1.6). After
doing this, the two LFs are in perfect agreement. Note that \cite{delmoro13}
have also shown that $q_{24} \le -0.23$ gives an incomplete selection of RL
AGN as it misses evidence for radio jets.
 
\begin{table*}
\caption{RL AGN Luminosity Function and Evolution: ZDE maximum likelihood
  analysis best fits. We consider a LF of the type $\Phi(P) \propto
    P^{-{\gamma_1}}$ and an evolution peaking at $z_{\rm peak}$, which is
    $\propto (1+z)^{k_{\rm low}}$ and $\propto (1+z)^{k_{\rm high}}$ at low and
    high redshifts respectively (see eq. \ref{eq:ev}).  KS$_{\rm z}$ and
    KS$_{\rm P}$ are the KS $p$-values for the redshift and radio power
    distributions respectively.}
\begin{tabular}{llccrcrr}
Sample&$\gamma_1$&$k_{\rm low}$&$k_{\rm high}$&$z_{\rm peak}$&KS$_{\rm z}$&KS$_{\rm P}$\\
\hline
RL AGN & $1.44\pm0.04$& $2.2^{+1.8}_{-1.6}$ & $-3.9^{+0.7}_{-0.8}$ & $0.49\pm0.10$ & 0.331& 0.889 \\
RL AGN, no LSS & $1.45\pm0.05$& $1.4^{+2.0}_{-1.2}$ & $-3.7^{+0.8}_{-1.4}$ & $0.41\pm0.18$ & 0.524& 0.872 \\
\hline
\end{tabular}
\label{tabmaxl_RLAGN}
\end{table*}

We also include the $0.05 < z < 0.35$ VIDEO-{\it XXM3} LF of \cite{McAl13}
(blue open squares), who selected their AGN based on them being in host galaxies
``redder than the spiral galaxy templates" (and therefore are most likely
going to be of the RL type). Their LF appears to be somewhat above ours at $P
\approx 10^{23}$ W Hz$^{-1}$ but the difference in selection criteria between
the two derivations makes a more detailed comparison difficult.

Finally, Fig. \ref{rl_agn_local} shows also two more LFs: the $z < 0.3$ LF of
\cite{best14} for all sources (red short-long dashed line) and radiative-mode
only (green dot - short-dashed line). Our local LF is consistent with the first LF
although nominally a factor $\sim 1.3$ times higher. We discuss this, together with the
general issue of comparing our LFs to many previous derivations, in
Sect. \ref{sect:rlagn_evol}. Our LF is $\sim 2$ orders of magnitude above
the radiative-mode radio AGN LF, which confirms the fact that the bulk of our
RL AGN are of the jet-mode type (see also Sect. \ref{astro}).

Fig. \ref{rl_agn_local} shows also two further estimates of the local LF for
E-CDFS RL AGN, which are based on the whole sample but are model
dependent. The first one is the LF de-evolved to $z=0$ (filled circles) using
the best fit evolutionary parameter for PDE from the $\langle V_{\rm
  e}/V_{\rm a} \rangle$ analysis. This LF is totally inconsistent ($p$-value $< 0.0001$) 
  with both previous LFs, which indicates that the redshift
evolution is more complex than assumed. (The situation does not change taking
into account the LSS: $p$-value $< 0.0001$). This is confirmed by the second
one, that is the LF derived from the maximum likelihood analysis using a more
complex density evolutionary law than PDE, as detailed in Sect.
\ref{rlagn:LF_z}. In this case the best fit LF is in good agreement with both
previous LFs up to $P \ga 10^{24}$ W Hz$^{-1}$, which suggests that this
evolutionary law might be valid only up to these powers, as we cannot yet
constrain the bright end slope of the LF 
(this has however been done by other wider/shallower surveys).

\subsubsection{The LF redshift evolution}\label{rlagn:LF_z}

We noticed in Sect. \ref{banded} that the RL AGN evolution cannot be simply
parametrized by a $(1+z)^k$ law but reaches a peak at $z \approx 0.5 - 1$. We
therefore introduce a more complex density evolutionary law, which we term as
``$z_{\rm peak}$ density evolution'' (ZDE) as follows:

\begin{equation}
ev(z) = \frac{(1+z_{\rm c})^{k_{\rm low}} + (1+z_{\rm c})]^{k_{\rm high}}}
  {[(1+z_{\rm c})/(1+z)]^{k_{\rm low}} + [(1+z_{\rm c})/(1+z)]^{k_{\rm
        high}}}.
\label{eq:ev}        
\end{equation}

\begin{figure}
\includegraphics[height=8.6cm]{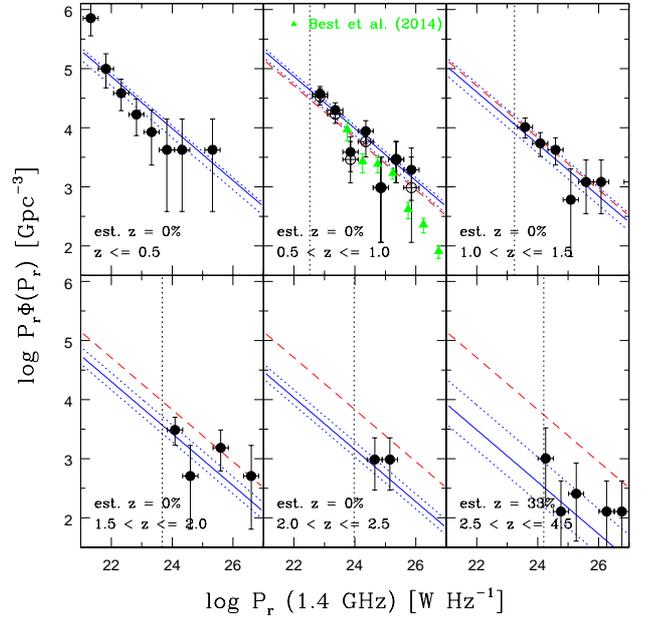}
\caption{The differential 1.4 GHz LF for RL AGN in a $P \times \phi(P)$ form in
  six redshift bins. The (blue) solid lines represent the best fit single
  power-law LF from the maximum likelihood method (ZDE) evolved to the central
  redshift of the bin using the best fit for the complex density evolution
  described above, with dotted lines showing the same LF at the two extreme
  redshifts defining the bin. The (red) short-dashed line represents the best
  fit LF at $z=0$. Error bars correspond to $1\sigma$ Poisson errors evaluated
  using the number of sources per bin with redshift determination added in
    quadrature with the cosmic variance uncertainties. The percentage of
  redshifts estimated from $f_{3.6\mu m}$ is also given for each bin. Open
  symbols in the $z = 0.5 - 1.0$ bin take into account the presence of LSS,
  while (green) triangles are the LF from \protect\cite{best14} (all
  sources). The dotted vertical lines denote the completeness limit in power for
  each redshift range. See text for details.}
\label{rlagn_z_bins}
\end{figure}

This allows for a situation where evolution changes sign above some redshift and
for a smooth transition. Indeed, $ev(z) \propto (1+z)^{k_{\rm low}}$ and $ev(z)
\propto (1+z)^{k_{\rm high}}$ for $z \ll z_{\rm c}$ and $z \gg z_{\rm c}$
respectively (and $ev(0) = 1$). If $k_{\rm low} > 0$ and $k_{\rm high} < 0$ then
$ev(z)$ peaks at $z_{\rm peak} = (-k_{\rm high}/k_{\rm low})^{1/(k_{\rm
    high}-k_{\rm low})}(1+z_{\rm c})-1$.  The results of fitting this model to
our RL AGN sample are reported in Table \ref{tabmaxl_RLAGN}. We find a best fit
$z_{\rm peak} \sim 0.5$, with the density evolution turning from positive to
negative respectively below and above this value, while the slope of the LF is
basically the same as derived before. Importantly, the model is consistent with
the data as shown by the results of both KS tests. Taking into account the
effects of the LSS changes our best fit parameters by less than $1 \sigma$.

Fig. \ref{rlagn_z_bins} (tabulated in Table \ref{tab:rlagn_lf_z}) shows the
evolution of the RL AGN LF in six redshift bins. The decrease in number at
higher redshifts is clearly visible, as is the fact that the ZDE maximum
likelihood fit provides a very good representation of the LF redshift
evolution.

\begin{table}
\caption{Luminosity Functions of VLA-ECDFS RL AGN.}
\begin{tabular}{lclr}
$z$ range & $\log P_{\rm 1.4GHz}$ &$\log P \Phi(P)$& ~~N\\
& {W~Hz$^{-1}$} & {Gpc$^{-3}$}& \\
\hline
  & 21.33 & $5.85^{+0.26}_{-0.30}$ &    4 \\
  & 21.83 & $4.99^{+0.26}_{-0.32}$ &    4 \\
  & 22.33 & $4.58^{+0.23}_{-0.30}$ &    5 \\
$ z \le 0.5$   & 22.83 & $4.22^{+0.26}_{-0.35}$ &    4 \\
  & 23.33 & $3.93^{+0.37}_{-0.56}$ &    2 \\
  & 23.83 & $3.62^{+0.52}_{-1.04}$ &    1 \\
  & 24.33 & $3.62^{+0.52}_{-1.04}$ &    1 \\
  & 25.33 & $3.62^{+0.52}_{-1.04}$ &    1 \\
 \hline
   & 22.85 & $4.57^{+0.13}_{-0.14}$ ($4.53^{+0.13}_{-0.15}$)&   18 (16)\\
  & 23.35 & $4.30^{+0.12}_{-0.14}$ ($4.23^{+0.13}_{-0.16}$) &   20 (17)\\
  & 23.85 & $3.59^{+0.26}_{-0.33}$ ($3.46^{+0.30}_{-0.39}$) &    4 (3)\\
$0.5< z \le 1.0$   & 24.35 & $3.94^{+0.18}_{-0.21}$ ($3.76^{+0.21}_{-0.26}$) &    9 (6)\\ 
  & 24.85 & $2.99^{+0.52}_{-0.93}$ &    1 \\
  & 25.35 & $3.46^{+0.30}_{-0.39}$ &    3 \\
  & 25.85 & $3.29^{+0.37}_{-0.52}$ ($2.99^{+0.52}_{-0.93}$) &    2 (1)\\
   \hline
  & 23.59 & $4.01^{+0.15}_{-0.18}$ &   13 \\
  & 24.09 & $3.74^{+0.18}_{-0.22}$ &    9 \\
$1.0 < z \le 1.5$   & 24.59 & $3.63^{+0.20}_{-0.25}$ &    7 \\
  & 25.09 & $2.78^{+0.52}_{-0.96}$ &    1 \\
  & 25.59 & $3.08^{+0.37}_{-0.54}$ &    2 \\
  & 26.09 & $3.08^{+0.37}_{-0.54}$ &    2 \\   
   \hline
  & 24.09 & $3.49^{+0.21}_{-0.26}$ &    6 \\
$1.5 < z \le 2.0$  & 24.59 & $2.71^{+0.52}_{-0.90}$ &    1 \\
  & 25.59 & $3.18^{+0.30}_{-0.39}$ &    3 \\
  & 26.59 & $2.71^{+0.52}_{-0.90}$ &    1 \\
 \hline
  & 24.65 & $2.98^{+0.37}_{-0.52}$ &    2 \\
$2.0 < z \le 2.5$   & 25.15 & $2.98^{+0.37}_{-0.52}$ &    2 \\
 \hline
  & 24.26 & $3.00^{+0.52}_{-0.79}$ &    1 \\
  & 24.76 & $2.11^{+0.52}_{-0.80}$ &    1 \\
$ 2.5 < z \le 4.5$  & 25.26 & $2.41^{+0.52}_{-0.80}$ &    2 \\
  & 26.26 & $2.11^{+0.52}_{-0.80}$ &    1 \\
  & 26.76 & $2.11^{+0.52}_{-0.80}$ &    1 \\
\hline
\multicolumn{4}{l}{\footnotesize Numbers in parenthesis take into account the effect of the}\\ 
\multicolumn{4}{l}{\footnotesize  LSS. Errors correspond
  to $1\sigma$ Poisson errors evaluated using}\\
\multicolumn{4}{l}{\footnotesize  the number of
  sources per bin with redshift determination}\\
\multicolumn{4}{l}{\footnotesize added in quadrature with the cosmic variance uncertainties.}\\ 
\multicolumn{4}{l}{\footnotesize The conversion to units of Mpc$^{-3}$ dex$^{-1}$ used, for example, by}\\
\multicolumn{4}{l}{\footnotesize \cite{Sim12}, is done by subtracting $9 - \log(\ln(10))$}\\
\multicolumn{4}{l}{\footnotesize from our values.}\\
\end{tabular}
\label{tab:rlagn_lf_z}
\end{table}

As discussed in Sect. \ref{sect:rlagn_evol}, a direct comparison between the
evolution of our LF and previously published results is not straightforward.
However, once one takes into account some of the most obvious differences our
results are in agreement with previous ones. As an example,
Fig. \ref{rlagn_z_bins} shows that our LF in the $0.5 - 1.0$ redshift bin is
somewhat higher, as expected, but still consistent within the errors, with that
of \cite{best14} (all sources). The same thing applies to the LFs of
\cite{gen10} in the $0.8 - 1.5$ and $1.2 - 2.5$ bins.

\subsection{RQ AGN}

\subsubsection{The local LF}\label{rqagn:LF_local}

\begin{figure}
\includegraphics[height=8.6cm]{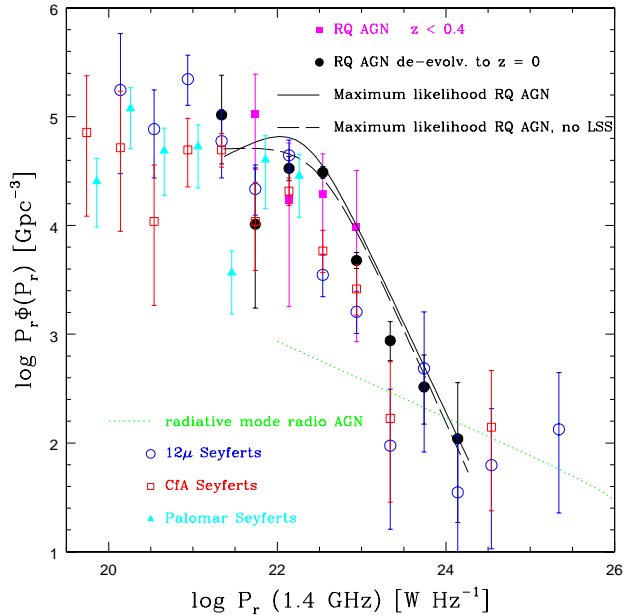}
\caption{The local differential 1.4 GHz LF for RQ AGN in a $P \times \phi(P)$
  form. Filled (magenta) squares triangles indicate the E-CDFS LF at $z \le
  0.4$, while filled (black) circles denote the LF de-evolved to $z=0$ using the
  best fit evolutionary parameter from the $\langle V_{\rm e}/V_{\rm a} \rangle$
  analysis. The solid line is the best-fit double power-law LF from the maximum
  likelihood method (PLE), while the dashed line is the same LF but taking into
  account the presence of LSS. The (green) dotted line is the best fit to the
  local LF of radiative-mode radio AGN \protect\citep{best14}. Local bivariate
  LFs for Seyfert galaxies from three samples are also shown: the $12\mu$m (open
  blue circles) and CfA (open red squares) Seyferts \citep{rush96} and the
  Palomar Seyferts \citep[filled cyan triangles;][]{ulv2001}. Error bars for the
  E-CDFS $z \le 0.4$ sample correspond to $1\sigma$ Poisson errors evaluated
  using the number of sources per bin with redshift determination added in
    quadrature with the cosmic variance uncertainties.}
\label{rq_agn_local}
\end{figure}

The local ($z \la 0.3 - 0.4$) LF for radio-selected RQ AGN has, so far, never
been determined\footnote{In \cite{pad11b} we estimated it by de-evolving the
  full LF to $z=0$.}.  We present it in Fig. \ref{rq_agn_local}. The RQ AGN $z
\le 0.4$ sample (filled squares) includes only 6 objects and the corresponding
LF is therefore quite uncertain. We therefore complement it by deriving the LF
for the full sample de-evolved to $z=0$ (filled circles) using the best fit
evolutionary parameter from the $\langle V_{\rm e}/V_{\rm a} \rangle$ analysis
and the local LF derived from the maximum likelihood analysis (2 power-law,
PLE).  Given the discussion in Sect. \ref{evolution} both these estimates should
be reliable. We also plot in Fig.  \ref{rq_agn_local} the radio LFs for the
$12\mu$m (open circles) and Center for Astrophysics (CfA; open squares) Seyferts
\citep{rush96} and the Palomar Seyferts \citep{ulv2001}, that is of the local
counterparts of (mostly) RQ AGN. There are a couple of points one needs to
consider here: 1) all these LFs are bivariate, that is they were derived by
observing in the radio band samples of Seyfert galaxies. Given that the original
sample selection was done in the optical, it is almost certain that the samples
are not complete in the radio and therefore the LFs are lower limits to the true
ones; 2) there is some scatter between the various estimates, although this is
usually within the rather large error bars, which are due to the fact that many
bins include only $1 - 2$ sources; 3) the high-power ($P \ga 10^{24}$ W
Hz$^{-1}$) tail should not be considered since the $12\mu$m sample, for example,
includes a number of ``classical'' RL sources (e.g. 3C 273, 3C 120, OJ
287). Taking all of the above into account, our determination of the local LF
for radio-selected RQ AGN is not inconsistent with the Seyfert LFs. We will
return to this issue in Sect. \ref{sect:rqagn_evol}.

\subsubsection{The LF redshift evolution}\label{rqagn:LF_z}

\begin{table}
\caption{Luminosity Functions of VLA-ECDFS RQ AGN.}
\begin{tabular}{lclr}
$z$ range & $\log P_{\rm 1.4GHz}$ &$\log P \Phi(P)$& ~~N\\
& {W~Hz$^{-1}$} & {Gpc$^{-3}$}& \\
\hline
  & 21.77 & $5.02^{+0.37}_{-0.49}$ &    2 \\
$ z \le 0.5$  & 22.17 & $4.24^{+0.52}_{-0.91}$ &    1 \\
  & 22.57 & $4.43^{+0.24}_{-0.31}$ &    5 \\
  & 22.97 & $4.21^{+0.37}_{-0.53}$ &    3 \\
\hline
  & 22.88 & $4.85^{+0.12}_{-0.14}$ ($4.48^{+0.16}_{-0.19}$) &   18 (10)\\
$ 0.5 < z \le 1.0$  & 23.28 & $4.46^{+0.12}_{-0.14}$ ($4.43^{+0.12}_{-0.14}$) &   20 (19)\\
  & 24.08 & $3.09^{+0.52}_{-0.91}$ &    1 \\
  & 24.48 & $3.09^{+0.52}_{-0.91}$ &    1 \\
  \hline
  & 23.50 & $4.65^{+0.13}_{-0.15}$ &   18 \\
$ 1.0 < z \le 1.5$   & 23.90 & $4.03^{+0.15}_{-0.18}$ &   14 \\
  & 24.30 & $3.59^{+0.26}_{-0.33}$ &    5 \\
 \hline
   & 23.57 & $3.53^{+0.37}_{-0.49}$ &    2 \\
$ 1.5 < z \le 2.0$ & 23.97 & $4.06^{+0.14}_{-0.16}$ ($4.03^{+0.15}_{-0.17}$)&   14 (13)\\
  & 24.37 & $3.71^{+0.18}_{-0.22}$ ($3.66^{+0.20}_{-0.24}$)&    8 (7)\\
  & 24.77 & $3.11^{+0.37}_{-0.51}$ ($2.81^{+0.52}_{-0.89}$) &    2 (1)\\
  \hline
   & 24.29 & $3.97^{+0.16}_{-0.19}$ &   11 \\
$ 2.0 < z \le 2.5$  & 24.69 & $3.09^{+0.37}_{-0.52}$ &    2 \\
  & 25.09 & $2.79^{+0.52}_{-0.91}$ &    1 \\
  & 25.49 & $2.79^{+0.52}_{-0.91}$ &    1 \\
  \hline
   & 24.32 & $3.75^{+0.18}_{-0.20}$ &    9 \\
$ 2.5 < z \le 3.66$   & 24.72 & $3.63^{+0.15}_{-0.17}$ &   14 \\
  & 25.12 & $2.73^{+0.37}_{-0.48}$ &    2 \\
  & 25.52 & $2.43^{+0.52}_{-0.83}$ &    1 \\  
\hline
\multicolumn{4}{l}{\footnotesize Numbers in parenthesis take into account the effect of the}\\ 
\multicolumn{4}{l}{\footnotesize  LSS. Errors correspond
  to $1\sigma$ Poisson errors evaluated using}\\
\multicolumn{4}{l}{\footnotesize  the number of
  sources per bin with redshift determination}\\
\multicolumn{4}{l}{\footnotesize added in quadrature with the cosmic variance uncertainties.}\\ 
\multicolumn{4}{l}{\footnotesize The conversion to units of Mpc$^{-3}$ dex$^{-1}$ used, for example, by}\\
\multicolumn{4}{l}{\footnotesize \cite{Sim12}, is done by subtracting $9 - \log(\ln(10))$}\\
\multicolumn{4}{l}{\footnotesize from our values.}\\
\end{tabular}
\label{tab:rqagn_lf_z}
\end{table}

\begin{figure}
\includegraphics[height=8.6cm]{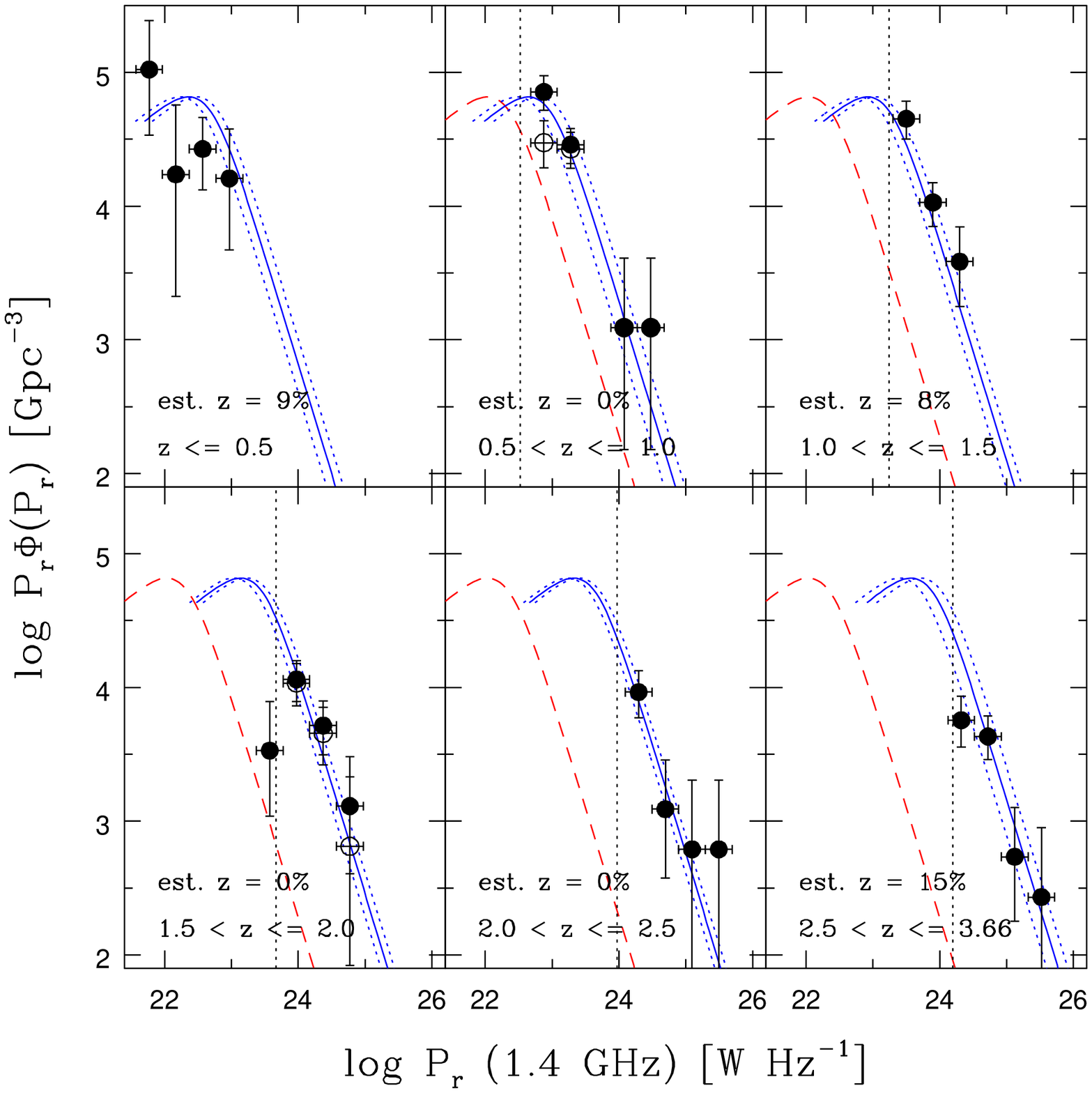}
\caption{The differential 1.4 GHz LF for RQ AGN in a $P \times \phi(P)$ form in
  six redshift bins. The (blue) solid lines represent the best fit double
  power-law LF from the maximum likelihood method evolved to the central
  redshift of the bin using the best fit for pure luminosity evolution
  $(1+z)^{2.52}$, with dotted lines showing the same LF at the two extreme
  redshifts defining the bin. The (red) short-dashed line represents the best
  fit LF at $z=0$. Error bars correspond to $1\sigma$ Poisson errors evaluated
  using the number of sources per bin with redshift determination added in
    quadrature with the cosmic variance uncertainties. The percentage of
  redshifts estimated from $f_{3.6\mu m}$ is also given for each bin. Open
  symbols in the $z = 0.5 - 1.0$ and $1.5 - 2.0$ bins take into account the
  presence of LSS. The (black) dotted vertical lines denote the completeness
  limit in power for each redshift range. See text for details.}
\label{rqagn_z_bins}
\end{figure}

The first ever determined evolution of the RQ AGN radio LF is shown in six
redshift bins in Fig. \ref{rqagn_z_bins} (tabulated in Table
\ref{tab:rqagn_lf_z}). The increase in power at higher redshifts is clearly
visible, as is the fact that the PLE maximum likelihood fit provides a good
representation of the LF redshift evolution, which means that the possible
slowdown at higher redshifts hinted at in Sect. \ref{banded} has to be subtle.

\subsection{RL and RQ AGN comparison}\label{agn:LF_z}

\begin{figure}
\includegraphics[height=8.6cm]{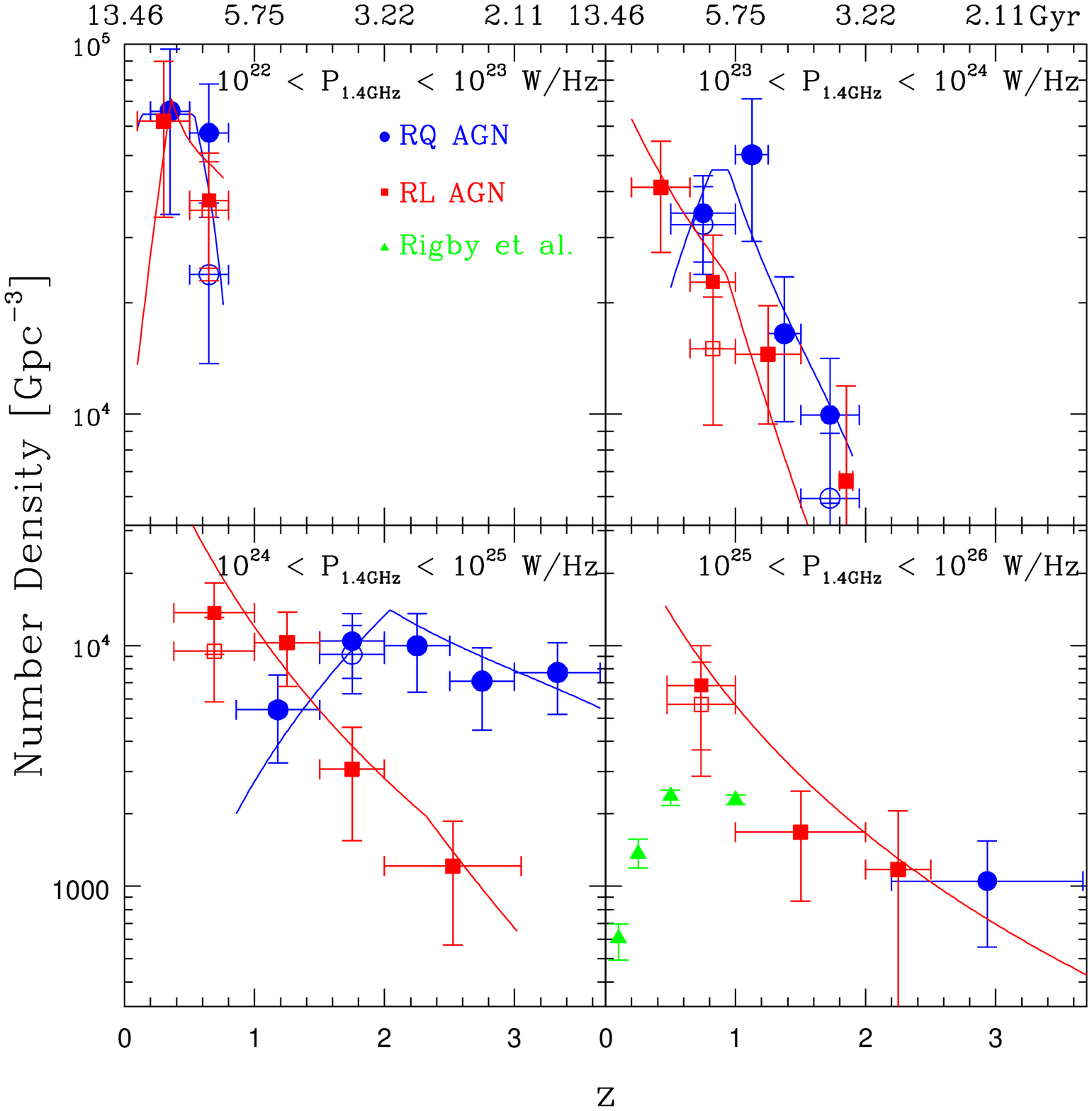}
\caption{The redshift evolution of RL (red squares) and RQ (blue circles) AGN
  in various luminosity bins. The top labels refer to cosmic time in Gyr.  The
  points come from the $1/V_{\rm a}$ method, while the lines are from the
  best fits assuming a single power law and PLE for RQ AGN and PDE for RL AGN
  respectively, taking into account the effect of our flux density
  limit. Error bars correspond to $1\sigma$ Poisson errors
  added in quadrature with the cosmic variance uncertainties. 
  Empty symbols refer to the ``no LSS'' samples. Green triangles are
  from \protect\cite{rig11}.}
\label{rq_rl_AGN_NT}
\end{figure}

Fig. \ref{rq_rl_AGN_NT} shows the redshift evolution of RL and RQ AGN in
various luminosity bins, with the points coming from the $1/V_{\rm a}$ method
and the lines being the best maximum likelihood fits assuming PLE for RQ AGN
and PDE for RL AGN respectively and a single power law (given the relatively
small power range). The figure shows that, while at low powers ($P_{\rm
  1.4GHz} < 10^{23}$ W Hz$^{-1}$) RQ and RL AGN are equally numerous, the
situation changes at $10^{23} < P_{\rm 1.4GHz} < 10^{25}$ W Hz$^{-1}$, where
RQ AGN overtake RL ones. For example, at $z \sim 2 - 3$ and $10^{24} < P_{\rm
  1.4GHz} < 10^{25}$ W Hz$^{-1}$ RQ AGN are more abundant than RL ones by
factors $\sim 3 - 10$. At even larger powers ($P_{\rm 1.4GHz} > 10^{25}$ W
Hz$^{-1}$) the number densities of the two classes appear to be similar again
but at this point we are running out of RQ AGN.

We stress that the fact that the decline appears to start at larger redshifts
for more powerful sources has nothing to do with ``cosmic down-sizing''
\citep[as instead stated by][]{Sim12} but is simply an effect of our flux density
limit. Within a redshift bin, in fact, not all sources in a given radio luminosity bin
might be visible by our survey and this leads to a ``loss'' of objects. As a result,
sources will start to fall below the flux density limit at
progressively higher redshifts for larger powers. This explains the sudden 
stop in the rise of both classes, which in deeper samples is expected to
continue (obviously barring an intrinsic redshift cutoff).

\cite{ue14} reach maximum number densities $\sim 3 \times 10^5$ Gpc$^{-3}$
for their X-ray selected (largely RQ) AGN with $10^{42} < L_{\rm x} <
10^{43}$ erg s$^{-1}$ and $z \sim 0.7$ (see their Fig. 12), to be compared
with our value of $\sim 7 \times 10^4$ Gpc$^{-3}$ at $z \sim 0.3$
(Fig. \ref{rq_rl_AGN_NT}). This is to be expected as their AGN surface
density is $\sim 3,000$ deg$^{-2}$ vs. $\sim 860$
deg$^{-2}$ for our RQ AGN (see Sect. \ref{sect:rqagn_evol}), which simply
means that we have not detected all AGN with $L_{\rm x} > 10^{42}$ erg
s$^{-1}$ (but only those above our radio flux density limit).

Finally, we note that the number densities for ``classical'' RL quasars are
orders of magnitude smaller than those of our RL AGN \citep[$\la 1$
Gpc$^{-3}$ for $P_{\rm 2.7GHz} \ga 10^{27}$ W Hz$^{-1}$: see, e.g. Fig. 5
of][]{wall05}.

\section{Discussion}\label{sec_disc}

\subsection{The evolution of micro-Jy radio sources}

We now analyze in more detail the evolution of the two AGN classes and, when
applicable, compare it with previous results.

\subsubsection{RL AGN}\label{sect:rlagn_evol}

The evolution of RL AGN is not described by PLE or a simple 
PDE. Instead, it can be characterised as a density evolution
reaching a peak at $z \sim 0.5$ and then decreasing.

\cite{rig11} studied various samples of {\it steep-spectrum} ($\alpha_{\rm r} >
0.5$) AGN selected from a variety of radio surveys with increasingly smaller
areas and flux density limits. They found a significant decline in comoving
density at $z > 0.7$ for their lower luminosity sources ($10^{25} < P_{\rm
  1.4GHz} < 10^{26}$ W Hz$^{-1}$). Turnovers were still present at higher powers
but moved to larger redshifts, suggesting a luminosity-dependent evolution of
the redshift peak, similar to the so-called ``cosmic downsizing'' seen, for
example, in the X-ray band \citep{ue14}. Fig. \ref{rq_rl_AGN_NT} (bottom right)
shows the relevant points (filled triangles) from Fig. 11 of \cite{rig11} in one
of the two ranges of luminosities we have in common. Considering that, unlike
these authors, we have no cut on $\alpha_{\rm r}$ and therefore we are bound to
have more AGN in our sample, our points agree with theirs within our error
bars. The same is true for the $10^{26} - 10^{27}$ W Hz$^{-1}$ range (not shown
because there are no RQ AGN at those powers). Fig. \ref{rq_rl_AGN_NT} shows that
at the powers corresponding to the low luminosity end of the \cite{rig11} sample
we cannot put any constraints on a peak at $z \sim 0.7$, as we are only
sensitive to $z \ga 0.5$, but our data are consistent with the trend present in
the \cite{rig11} sample. Moreover, we can use Fig. 10 of \cite{rig11}, which
shows a strong correlation between $z_{\rm peak}$ and power. Since our median
$P_{\rm 1.4GHz}$ is $\sim 10^{24}$ W Hz$^{-1}$ the $z_{\rm peak}$ $\sim 0.5$ we
find through the ZDE fit appears to be in agreement with the \cite{rig11}
results.

\cite{best14} have also studied various samples of (steep-spectrum) radio AGN
selected from a variety of surveys with increasingly smaller areas and flux
density limits. They classify their sources into radiative-mode and jet-mode
AGN using emission line diagnostics. Based on Fig. \ref{rl_agn_local},
\cite{bon13}, and Sect. \ref{astro} \citep[see also Fig. 1 of][]{best14}, the
large majority of our RL AGN have to be of the latter type, so we compare our
findings to their results on the jet-mode population.  For $P_{\rm 1.4GHz}
\la 10^{24}$ W Hz$^{-1}$ \cite{best14} (see their Fig. 5) find that its space
density stays constant up to $z \approx 0.5$ and then decreases to $z =
1$. At moderate powers, $10^{24} \la P_{\rm 1.4GHz} \la 10^{26}$ W Hz$^{-1}$,
the space density increases to $z \sim 0.5$ before falling. At the highest
powers the space density appears to increase up to $z \sim 1$ but the
statistics is somewhat limited.

Our results are similar to those of \cite{best14}, taking into account our
somewhat more limited coverage of the luminosity -- redshift
plane. Fig. \ref{rq_rl_AGN_NT} shows that for $P_{\rm 1.4GHz} \la 10^{23}$ W
Hz$^{-1}$ the space density of RL AGN increases up to $z \approx 0.5$ while for
$10^{24} \la P_{\rm 1.4GHz} \la 10^{26}$ W Hz$^{-1}$, and taking into account
the effect of the LSS, it increases to $z \approx 1.0$ before falling.  We
cannot say anything about larger powers.
 
It is important to stress the difference between a ``purely flux density
limited'' sample, like ours, and many others published in the literature, which
are very often put together by cross-correlating very large surveys. For
example, the \cite{best14} LF, as discussed in \cite{hec14}, is built by 
combining four different radio LFs based on four spectroscopic samples:
Las Campanas \citep{mac00}, 2dFGRS \citep{sad02}, SDSS/NVSS/FIRST
\citep{best12}, and 6dFGS \citep{mau07}. Although these samples are large
(including, e.g. more than 2,500 and 7,000 sources for the 6dFGS and
SDSS/NVSS/FIRST samples) one needs to consider that, by construction, they do
not sample the full RL AGN population. All of them, in fact, are limited to
galaxies (and therefore do not include non-stellar and broad-line sources) and
to steep-spectrum ($\alpha_{\rm r} > 0.5$) radio sources. \cite{gen10} and
\cite{gen13}, who derived the LF for FR I/II radio galaxies, also included only
galaxies, excluding any compact or quasar-dominated source.
Our LF, being simply radio flux density limited, has no such biases and therefore is
bound to find number densities larger than these determinations; this is true
also for \cite{Mao12} and \cite{Sim12} (pending their different
definition of RL AGN).

\cite{pad11b} made a connection between the negative evolution of
RL AGN and that of elliptical galaxies. For example, \cite{tay09} have found
that the number density of massive ($M_{\star} > 10^{11} M_{\odot}$) red
galaxies declines with redshift as $\Phi(z) \propto
(1+z)^{-1.60\pm0.14(\pm0.21)}$ for $z \le 1.8${\footnote{\cite{best14} have 
compiled some more recent results, which paint a somewhat more complex evolution.}. 
If we restrict our sample to
$z \le 1.8$ we get $\Phi(z) \propto (1+z)^{-1.3\pm0.5}$, which is very
similar. 

Fig. \ref{Best_z_1p5} compares the LF of our RL AGN in the redshift
range $1.3 - 1.7$
with two different models at $z = 1.5$ from \cite{best14}. These models,
starting from the assumption that jet-mode RL AGN are hosted in quiescent
galaxies, combine the known stellar mass function of the host galaxies with the
stellar mass dependency of the presence of a RL AGN to make predictions on
the evolution of the RL AGN LF. Model 2b is a luminosity-density evolution
model, in which the luminosity of the RL AGN increases with increasing redshift
and there is a delay of $\sim 2$ Gyr between the formation of the host galaxy
and the presence of a radio jet. Model 3a is a pure density evolution model in
which a subset of the sources classified as jet-mode are in reality related to
the radiative-mode RL AGN population.  This could be due to a misclassification
of some objects or to a change in physical properties with redshift. As shown in
Fig. 7 of \cite{best14}, the LFs predicted by these models are clearly
distinguishable only at $z>1$, i.e. beyond their maximum redshift.  Our sample
covers a wide radio power range also at $z \sim 1.5$ and therefore we have the
opportunity to test these different models. As shown in Fig. \ref{Best_z_1p5},
model 3a is consistent with our LF, while model 2b is clearly not.  Since the
predictions were based on jet-mode only RL AGN, we also restricted our sample
only to this type of sources by following \cite{bon13} and using the 22 $\mu$m
power criterion proposed by \cite{gur14}\footnote{As discussed in \cite{bon13},
  this method has not been tested on as faint a sample as ours and therefore we
  regard this as only a rough classification.}. Only one of our jet-mode RL AGN
turned out to have an Eddington ratio inconsistent with this classification ($>
0.1$; see Sect. \ref{astro}) and was therefore re-classified. The jet-mode LF is
shown in Fig. \ref{Best_z_1p5} (filled red triangles) and is basically the same
as the overall LF.

\begin{figure}
\includegraphics[height=8.6cm]{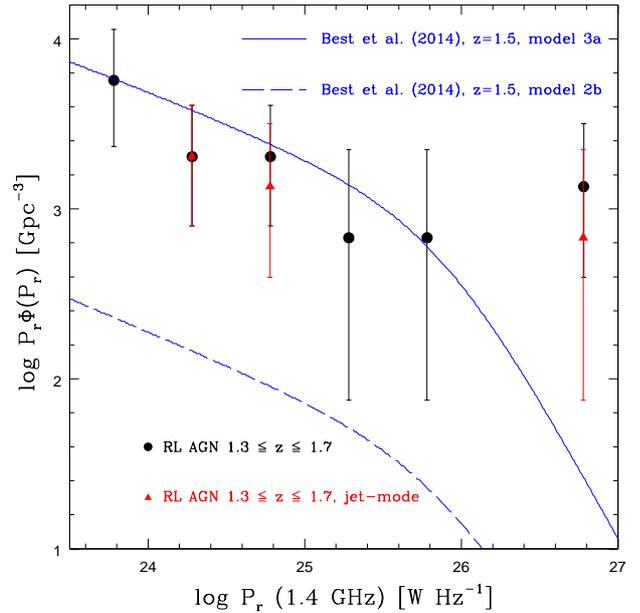}
\caption{The RL AGN LF at $1.3 \le z \le 1.7$ compared with two different models
  from \protect\cite{best14} at $z=1.5$. Filled (black) circles include all
  sources, while (red) triangles represent our best estimate of the jet-mode
  LF. See text for details.}
\label{Best_z_1p5}
\end{figure}

\subsubsection{RQ AGN}\label{sect:rqagn_evol}

We have provided a more robust estimate of the evolution of RQ AGN in the
radio band, following upon the first one derived in \cite{pad11b}. We have
modelled it as a pure luminosity evolution $P(z) \propto (1+z)^{k_L}$,
obtaining $k_L = 3.0\pm0.2$ from the $V_{\rm e}/V_{\rm a}$ analysis and $k_L
= 2.5\pm0.2$ from the maximum likelihood one, in both cases over the $0.2 -
3.66$ redshift range.

The redshift evolution of X-ray selected (largely RQ) AGN is best described by a
more complex evolutionary model, the so-called luminosity-dependent density
evolution (LDDE), which takes into account the observed steepening with redshift
in the faint-end slope of the LF and the fact that luminosity evolution appears
to stop at $z \approx 2.5$, as shown, for example, in Fig. 11 of \cite{ue14}. It
is important to note that the latter LF, and all those studied in the X-ray
band, refer only to Compton thin AGN, characterised by an absorbing column
density $N_{\rm H} < 1.5 \times 10^{24}$ cm$^{-2}$, while in principle we have
no such limitation (although our cut in X-ray power might also exclude Compton
thick sources). Keeping that in mind, Fig. \ref{rqagn_z_bins} shows that the LF
faint-end slope is constrained only at $z < 0.5$; moreover, most ($\sim 74\%$)
of our RQ AGN have $ z < 2$. The two effects combined make any deviation from
PLE of the type seen in the X-ray band hard to detect.

We nevertheless split the RQ AGN sample in two sub-samples of the same size with
redshift respectively smaller and larger than $1.3$, fitting PLE evolution to
both. Evolution slows down at higher redshifts, with $k = 4.0 \pm 0.6$ and $k =
2.0 \pm 0.5$ respectively below and above $z = 1.3$. However, the errors on the
evolutionary parameters overlap at the $\sim 2 \sigma$ level, making this
difference not statistically significant. In other words, a high-redshift
slowing down of the evolution is consistent with, but not required by, our data.

We note that the surface density of our RQ AGN, $\sim 860$ deg$^{-2}$, is
already larger than that reached by the deepest optical surveys, that is
$\sim 670$ deg$^{-2}$ for type I and II combined. This number is the sum of
$\sim 200$ deg$^{-2}$, from a type II AGN sample based on the zCOSMOS survey
and selected on the basis of the optical emission line ratios \citep{Bongio10},
and $\sim 470$ deg$^{-2}$, from a type I AGN sample extracted from the VIMOS
VLT Deep Survey \citep{Gav06}. We note that the overall (RQ plus RL) AGN surface 
density is $\sim 1,500$ deg$^{-2}$. 

As regards the X-ray band, 
our RQ AGN surface density is $\sim 20$ times smaller than that of
\cite{Leh12}, which is simply due to the fact that our radio flux density
limit is still not as deep as the equivalent one in the X-rays \citep[see
also Fig. 10 in][for a comparison of the star formation rates reachable in
the two bands]{vat12}.

To allow for a fairer comparison\footnote{Some of the topics discussed here
  and in Sect. \ref{fraction_RL} were originally presented in
  \cite{pad11a}. We give here updated numbers.} between our sample and the
X-ray samples we have fitted a double power law LF to all points in
Fig. \ref{rq_agn_local}, obtaining $N_{\rm T} \sim 2.7 \times 10^5$
Gpc$^{-3}$, a value, which is basically the same as the maximum number
density of \cite{ue14}. This implies a surface density $\sim 15,400$
deg$^{-2}$ \citep[using eq. 2 of][]{pad11a}, to be compared with $\sim
15,000$ deg$^{-2}$ in the 4 Ms survey of \cite{Leh12}. In other words, once
one takes into account the fact that our radio flux density limit is still
not as deep as the equivalent one in the X-ray band by including the radio LF
of Seyfert galaxies, the number and surface densities one obtains are the
same, which shows that we are simply looking at the same population from two
different bands and the sources we are selecting in the radio band are the
same as the X-ray emitting AGN.

\subsection{Astrophysics of micro-Jy sources}\label{astro}

\begin{figure}
\includegraphics[height=8.6cm]{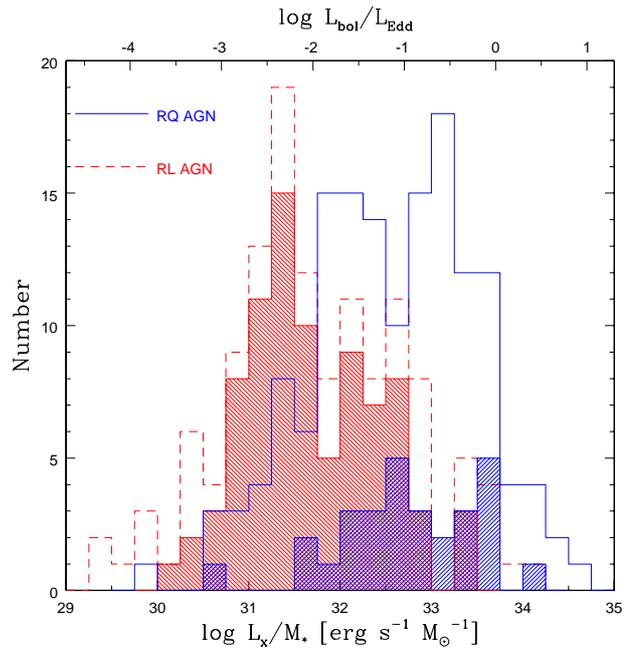}
\caption{The distribution of the ratio of X-ray power to stellar mass,
  $L_{\rm x}/M_{\star}$ for RQ (blue solid histogram) and RL (red dashed
  histogram) AGN. The top labels give the approximate values of the Eddington
  ratio. Shaded areas are upper limits.}
\label{Lx_M}
\end{figure}

To analyse in more depth the astrophysical nature of our AGN, we study the
ratio of X-ray power to stellar mass, $L_{\rm x}/M_{\star}$, a proxy for the
Eddington ratio $L/L_{\rm Edd}$ \citep[see, e.g.][]{Bongio12}. This can in
fact be written as

\begin{equation}
\frac{L}{L_{\rm Edd}} = \frac{k_{\rm bol}}{1.26 \times 10^{38} \Upsilon} \times 
\frac{L_{\rm x}}{M_{\star}},
\end{equation}
where $k_{\rm bol}$ is the $2-10$ keV bolometric correction and $\Upsilon =
M_{\rm BH}/M_{\star}$ is the ratio between black hole and stellar mass.

Stellar masses were estimated by \cite{bon13}, using the method of
\cite{Bongio12}. Fig. \ref{Lx_M} displays the distribution of $L_{\rm
  x}/M_{\star}$ for RQ (blue solid histogram) and RL (red dashed histogram)
AGN, with shaded areas denoting upper limits. We assume $k_{\rm bol} = 25$
\citep{Bongio12} and $\Upsilon = 0.001$ \citep[][and references
therein]{trump15}, noting that, given the observational scatter and a
possible dependence of $k_{\rm bol}$ on $L/L_{\rm Edd}$, the resulting
$L/L_{\rm Edd}$ values are only approximate. These are given by the top
labels in Fig. \ref{Lx_M}. For our choice of parameters, $L_{\rm x}/M_{\star}
= 5 \times 10^{33}$ erg s$^{-1}$ $M_{\odot}^{-1}$ corresponds to the
Eddington limit.

Fig. \ref{Lx_M} shows that the two $L_{\rm x}/M_{\star}$ distributions appear
very different, with RQ AGN having much larger values. Given the substantial
fraction of upper limits (all on $L_{\rm x}$) we used ASURV \citep{la92}, the
Survival Analysis package which employs the routines described in \cite{fei85}
and \cite{iso86}, which evaluate mean values by dealing properly with
non-detections and also compute the probability that two samples are drawn from
the same parent population. RQ and RL AGN have totally different ($p$-values
$<0.001$) distributions, according to all two-sample tests in ASURV.  The means
are also very different, being $L_{\rm x}/M_{\star} \sim 2.4 \times 10^{32}$ erg
s$^{-1}$ $M_{\odot}^{-1}$ and $\sim 5.5 \times 10^{30}$ erg s$^{-1}$
$M_{\odot}^{-1}$ for RQ and RL AGN respectively, a factor $\sim 40$
difference. The corresponding (approximate) Eddington ratios are $\approx 0.05$
and $\approx 0.001$. Given that the dividing line between radiative- and
jet-mode sources is thought to be $L/L_{\rm Edd} \approx 0.01$
\citep[e.g.][]{hec14}, our results are consistent with our RQ and RL AGN being
mostly of the former and latter type, respectively.

\subsection{The intrinsic fraction of radiative-mode RL AGN}\label{fraction_RL}

The RL AGN fraction, $f_{\rm RLAGN}$, has been derived in the past from
bright, optically selected samples which contain RL sources very different
from most of the sources studied here. Namely, they are RL quasars and
therefore belong with the radiative-mode AGN, while we have mostly jet-mode
objects (Sect. \ref{astro}). Indeed, the RL AGN in the ``classic'' Palomar
Green sample studied by \cite{kel89}, from which the ``standard'' value
$f_{\rm RLAGN} \sim 10\%$ comes,\footnote{Although in that paper the quoted
  value is $15 - 20\%$.}  have $P_{\rm 5GHz} \ga 10^{25}$ W Hz$^{-1}$ and
maximum number densities $P \Phi(P) \sim 11$ Gpc$^{-3}$ \citep{pad93}, that
is they are below the plotting area in Figs. \ref{agn_local} and
\ref{rl_agn_local} and barely inside in Fig. \ref{rq_agn_local}.

The way this fraction was derived was by looking for a bimodality in the
radio-to-optical flux density ratio, $R$, classifying as RL the sources with $R$
larger than the value corresponding to the minimum in the distribution
\citep[$0.1 < R < 100$ in][]{kel89}\footnote{We obviously cannot study the $R$
  distribution to look for a possible bi-modality as our RL AGN are mostly
  jet-mode sources while our RQ AGN are mostly radiative-mode sources (see
  Sect. \ref{astro}). Furthermore, as discussed by \cite{pad09} and
  \cite{pad11a}, $R$ is useful for quasar samples, where it is related to the
  jet/disk ratio, but loses its meaning for radio galaxies.}. \cite{pad11a} has
pointed out that: a) bright, optically selected samples might include RL quasars
with their optical flux boosted by relativistic beaming \citep[e.g.][]{gol99},
which artificially increases their fraction; b) RL AGN are on average more
powerful than RQ ones in the optical band \citep[e.g.][]{zam08}. When one
reaches the very faint end of the optical LF only RQ AGN will be present and
therefore the integrated, intrinsic RL fraction will be quite small. Both of
these effects lead to an overestimate of the RL AGN fraction when using the
``standard'' method. Indeed, it has been known for some time that the radio-loud
fraction drops with decreasing optical luminosity \citep{pad93}.  \cite{ji07}
have also shown that the fraction of RL quasars at $z=0.5$ declines from 24\% to
6\% as luminosity decreases from $M_{2500} = -26$ to $-22$. In other words, the
usually quoted value refers to the bright part of the LF and, when integrated
over the full range of powers, the resulting $f_{\rm RLAGN}$ is much smaller.

The best way to approach the question of the {\it intrinsic} value of $f_{\rm
  RLAGN}$ is to look at the band under discussion, namely the radio one. We plot
in Fig. \ref{rq_agn_local} our best estimate(s) of the local LF for RQ
(radiative-mode) AGN, together with the $z < 0.3$ LF of radiative-mode RL AGN
from \cite{best14} (dotted line). One can see that the fraction of
radiative-mode RL AGN is a strong function of radio power, as the shape of the
two LFs is very different. In particular, for $P_{\rm 1.4GHz} \ga 10^{24}$ W
Hz$^{-1}$ radiative-mode RL AGN dominates and $f_{\rm RLAGN} \ga 0.5$, while for
$P_{\rm 1.4GHz} \ge 10^{22}$ W Hz$^{-1}$, which is the limiting value for
radiative-mode RL AGN, $f_{\rm RLAGN} \sim 0.04$ (where we have used the double
power law fit derived in the previous sub-section). To derive the overall
intrinsic value one would need to integrate down to lower values but we do not
yet have a handle on the low-end of the radiative-mode RL AGN LF.

Our RQ AGN include absorbed (type II) and unabsorbed (type I) sources, while
the LF of radiative-mode RL AGN from \cite{best14} is limited to radio
galaxies and steep-spectrum radio sources (Sect. \ref{sect:rlagn_evol}).
However, as radio quasars are relativistically beamed, they tend to be much
more luminous and less numerous than radio galaxies and therefore we believe
that down to $P_{\rm 1.4GHz} \sim 10^{22}$ W Hz$^{-1}$ their exclusion makes
very little difference.

We note that if one takes as dividing line between RQ and radiative-mode RL
AGN the power at which the two LFs cross, one gets $P_{\rm 1.4GHz} \approx
10^{24}$ W Hz$^{-1}$ at $z \sim 0$, i.e., the same dividing value between the
two classes that has been suggested in the past \citep[see Sect. 2
of][]{pad93}. Given the strong redshift evolution of both LFs, one can
predict that this power is redshift dependent and will get larger at higher
redshifts, as indeed appears to be the case \citep{pad93}. Future radio
studies of deeper/wider samples, which will include RL AGN of the jet-mode
and radiative-mode type, should be able to study the RL fraction dependence
on redshift by deriving the evolving LFs for both classes.

\section{Summary and Conclusions}

We have used a deep, complete radio sample of 680 objects down to a 1.4 GHz
flux density of $32.5~\mu$Jy selected in the E-CDFS area to derive the number
counts of the various sub-mJy classes and to study in detail the evolution
and luminosity functions of radio faint AGN up to $z \sim 4$. Our main
results can be summarized as follows:

\begin{enumerate}

\item What we consider to be the best determination of the source population
  of the sub-mJy radio sky shows that star-forming galaxies and AGN make up a
  roughly equal part of the sub-mJy sky down to $32.5~\mu$Jy, with the former
  becoming the dominant population only below $\sim 0.1$ mJy.  RQ AGN are
  confirmed to be an important class of sub-mJy sources, accounting for $\sim
  25\%$ of the sample and $\sim 60\%$ of all AGN, and outnumbering RL AGN at
  $\la 0.1$ mJy (Sect. \ref{sec_number_counts}).

\item The AGN that make up the faint radio sky consists of two totally
  distinct populations: RQ AGN, mostly of the radiative-mode type, and RL
  AGN, largely of the jet-mode kind, characterised by very different evolutions,
  LFs, and Eddington ratios (Sects. \ref{sect:LF} and \ref{sec_disc}).

\item RQ AGN evolve in radio power as $\proptoapprox (1+z)^{2.5}$, similarly
  to star-forming galaxies, up to $z \sim 4$ but with a hint of a slowing
  down above $z \sim 1.3$.  Their LF is steep ($\Phi(P) \propto P^{-2.5}$ for
  $P \ga 2 \times 10^{22}$ W Hz$^{-1}$) and their $L/L_{\rm Edd}$ is
  typically $\ga 0.01$ (Table \ref{tabmaxl} and Sects. \ref{sect:rqagn_evol}
  and \ref{astro}).

\item RL AGN evolve in number $\proptoapprox (1+z)^{2}$ but only up to $z
  \sim 0.5$, beyond which their number density declines steeply $\propto
  (1+z)^{-4}$. Their LF is flat ($\Phi(P) \propto P^{-1.4}$) and their
  $L/L_{\rm Edd}$ is typically $\la 0.01$ (Table \ref{tabmaxl_RLAGN} and
  Sects. \ref{sect:rlagn_evol} and \ref{astro}).

\item The first determination of the local radio LF of RQ AGN appears to be
  consistent with that of Seyferts. Putting the two together, one derives a
  number density of RQ AGN equivalent to that of X-ray selected AGN, which
  shows we are looking at the same population from two different bands
  (Sects. \ref{rqagn:LF_local} and \ref{sect:rqagn_evol}).

\item We have approached the question of the {\it intrinsic} value of the
  fraction of radiative-mode RL sources, $f_{\rm RLAGN}$, within the AGN
  population by using, for the first time, the radio band (while all previous
  attempts were normally done in the optical band). By combining the local
  radio LF of Seyferts and our RQ AGN with a previously published estimate of the
  LF of radiative-mode RL AGN, we find that for $P_{\rm 1.4GHz} \ga 10^{24}$
  W Hz$^{-1}$ RL AGN dominates and $f_{\rm RLAGN} \ga 0.5$, while down to
  $P_{\rm 1.4GHz} \sim 10^{22}$ W Hz$^{-1}$ $f_{\rm RLAGN} \sim 4\%$ (Sect.
  \ref{fraction_RL}).
  
\item The surface density of radio-selected, radio-quiet AGN, $\sim 860$
  deg$^{-2}$, is already larger than that reached by the deepest optical
  surveys. This means that sub-mJy radio surveys, given the appropriate
  ancillary multi-wavelength data, have already the potential of detecting
  large numbers of radio-quiet AGN by-passing the problems of obscuration,
  which plague the optical and soft X-ray bands
  (Sect. \ref{sect:rqagn_evol}).
 
 \end{enumerate}

 In an upcoming paper (Padovani et al., in preparation) we will discuss in
 detail the evolution and LF of the SFG in our sample and how they compare to
 those of RQ AGN, which is very relevant for the issue of the mechanism
 behind radio emission in RQ AGN (see also \citealt{pad11b} and Bonzini et al. 2015, 
 submitted).

\section*{Acknowledgments}
We thank Philip Best, Ed Fomalont, Rachel Somerville, Jasper Wall,
and an anonymous referee for helpful comments and discussions and Piero
Rosati and Shaji Vattakunnel for their work on the E-CDFS survey.  We
acknowledge the ESO/GOODS project for the ISAAC and FORS2 data obtained using
the Very Large Telescope at the ESO Paranal Observatory under Program ID(s):
LP168.A-0485, 170.A-0788, 074.A-0709, 275.A-5060, and 081.A-0525.  We made use
of the TOPCAT software package \citep{tay05}. The VLA is a facility of the
National Radio Astronomy Observatory which is operated by Associated
Universities, Inc., under a cooperative agreement with the National Science
Foundation.  This research has made use of NASA's Astrophysics Data System (ADS)
Bibliographic Services.

\appendix
\section{Redshift estimates}\label{sec:red_estim}

\begin{figure}
\includegraphics[height=8.6cm]{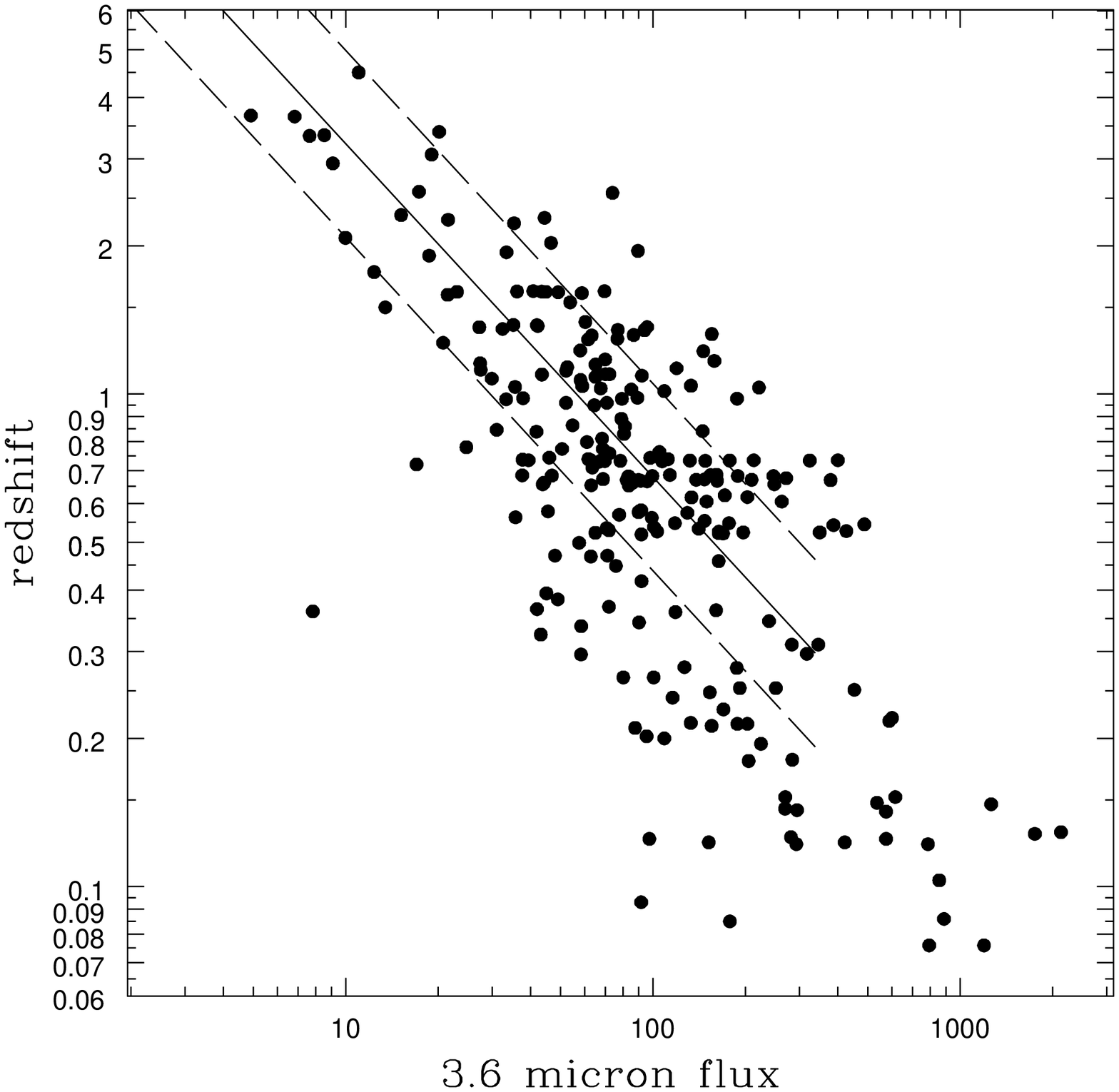}
\vspace{0.8cm}
\caption{Redshift versus $3.6\mu$m flux density for our sources with
  reasonable and secure redshift information \citep[quality flag $\ge 2$ as
  defined in][]{bon12}. The solid line is the best fit for $z \ge 0.3$, while
  the dashed lines represent the root mean square error. See text for
  details.}
\label{z3p6}
\end{figure}

As shown in Fig. \ref{z3p6}, reasonable and secure redshifts \citep[i.e.,
  spectroscopic with quality flag $\ge 2$, as defined
  in][]{bon12} are strongly correlated with $f_{3.6\mu m}$, albeit with
some scatter in particular at low redshift. The best and simplest approach
to estimate the redshift for the $\sim 8\%$ of the objects in the sample
without observed redshift is then to derive it from their $f_{3.6\mu m}$
by using the relationship shown in the figure (solid line), that
is $\log z = -0.677~log f_{3.6\mu m} + 1.185$. This was derived applying
the ordinary least-square bisector method \citep{iso90}, which treats the
variables symmetrically, to the $z \ge 0.3$ sample. The low redshift cut
was applied for three reasons: 1. in order to reduce the scatter in the
input data; 2. because the maximum $f_{3.6\mu m}$ of the sources without
redshift is $\sim 200$ microJy, which corresponds to $z \sim 0.3$ deriving
the best fit for the whole sample; 3. because the median $f_{3.6\mu m}$ for
sources without redshift is $\sim 10$ microJy and therefore they should
typically be at large ($z > 1$) redshift; inclusion of the low-redshift
sources in the sample might therefore bias our estimates. We note that the
best fits for the different sub-classes give redshift values, which differ
only by $\sim 0.4$ for sources having $f_{3.6\mu m} \sim 10$ microJy (the
median for sources without redshift). Note also that for $f_{3.6\mu m} =
1$, the minimum value for the sample, $z \sim 15.3$, which is on the high
side. We therefore set the maximum redshift equal to the maximum observed
values (spectroscopic or photometric) for the three sub-classes, namely $z
\sim 4.7$, 7.0, and 4.5 for SFG, radio-quiet, and radio-loud AGN
respectively.

\section{The large scale structure of the E-CDFS}\label{sec:LSS}

\cite{gil03} have studied the large-scale structure (LSS) in the CDFS in the
X-ray band, finding two main, narrow ($\Delta z \la 0.02$) structures at $ z
= 0.67$ and 0.73 and other spikes at 1.04, 1.62, and 2.57. Indeed, the
redshift distributions shown in Fig. \ref{histz} peak in the $0.5 - 0.75$ bin
for all classes. Given the small area of our survey one could worry that
such redshift spikes might influence some of our results. We then studied the
LSS of the E-CDFS by looking for narrow ($\Delta z \sim 0.01 - 0.03$)
structures and smoothing the observed spectroscopic redshift distribution in
large ($\Delta z \sim 0.1 - 0.3$) bins to estimate the background
distribution. Redshift peaks were defined as spikes being $> 2 \sigma$ above
the background for Poissonian errors. Applying this procedure to the whole
sample we recovered the main structures of \cite{gil03}, namely those at $z
\sim 0.67$, 0.73, and 1.62. We also find a structure at $z = 0.52$, present in
the near-IR band data of \cite{gil03}, and a new one at $z \sim
0.12$. Studying the LSS for the various sub-classes, however, showed that
not all redshift peaks are present for all classes. We then studied the LSS
individually finding the following significant peaks: 1) SFG: $0.122 - 0.126$
and $0.729 - 0.736$; AGN: $0.521 - 0.537$, $0.661 - 0.686$, $0.720 - 0.738$,
and $1.603 - 1.619$; RQ AGN: $0.521 - 0.529$, $0.652 - 0.671$, and $1.591 -
1.619$; RL AGN: $0.675 - 0.686$ and $0.731 - 0.734$. We then created ``no
LSS'' sub-samples by removing the sources in the bins listed above in excess
of the background distribution. The selection of the sources to be kept, which
typically included $\sim 20\%$ of those in the bin, was
done by picking the object(s) with radio flux densities closest to the median
value of the bin.

\label{lastpage}
\end{document}